\documentclass[journal]{IEEEtran} 
\IEEEoverridecommandlockouts

\usepackage[colorlinks]{hyperref}
\usepackage{mathrsfs}
\usepackage[noadjust]{cite}
\usepackage{graphicx,color,overpic,psfrag}
\usepackage{amsmath, amssymb}
\usepackage{latexsym}
\usepackage{bm}
\usepackage{amssymb}
\usepackage{cases}
\usepackage{array}
\usepackage{fancyhdr}
\usepackage{setspace}
\usepackage{caption}
\usepackage{indentfirst}
\usepackage{cases}
\usepackage{url}
\usepackage{algpseudocode}
\usepackage{algorithm}
\usepackage{blkarray}
\usepackage{subfig}
\usepackage{graphicx}

\usepackage{booktabs}
\usepackage{multirow}
\usepackage{dsfont}
\usepackage{tabularx}
\usepackage[table]{xcolor}
\usepackage{amsfonts}
\usepackage{amsthm}
\usepackage{stfloats}
\usepackage{letltxmacro}
\usepackage{lettrine}

\newcommand{\ba}{\mathbf{a}}

\newcommand{\bg}{\mathbf{g}}

\newcommand{\bp}{\mathbf{p}}

\newcommand{\bs}{\mathbf{s}}
\newcommand{\bt}{\mathbf{t}}

\newcommand{\bx}{\mathbf{x}}

\newcommand{\bB}{\mathbf{B}}

\newcommand{\bI}{\mathbf{I}}

\newcommand{\bP}{\mathbf{P}}

\newcommand{\bR}{\mathbf{R}}

\newcommand{\bW}{\mathbf{W}}

\graphicspath{{figure/}}
\allowdisplaybreaks

\begin{document}

\title{Specific Absorption Rate-Aware Multiuser MIMO Assisted by Fluid Antenna System}
\author{Yuqi~Ye, 
            Li~You,~\IEEEmembership{Senior~Member,~IEEE,} 
            Hao Xu,~\IEEEmembership{Senior~Member,~IEEE,} 
            Ahmed Elzanaty,~\IEEEmembership{Senior~Member,~IEEE,} 
            Kai-Kit Wong,~\IEEEmembership{Fellow,~IEEE,} 
            and Xiqi~Gao,~\IEEEmembership{Fellow,~IEEE}
\vspace{-5mm}		
		
\thanks{

Yuqi Ye, Li You, and Xiqi Gao are with the National Mobile Communications Research Laboratory, Southeast University, Nanjing 210096, China, and also with the Purple Mountain Laboratories, Nanjing 211100, China (e-mail: $\rm yqye@seu.edu.cn$; $\rm lyou@seu.edu.cn$; $\rm xqgao@seu.edu.cn$).

Hao Xu is with the National Mobile Communications Research Laboratory, Southeast University, Nanjing 210096, China (e-mail:$\rm hao.xu@seu.edu.cn$).

Kai-Kit Wong is affiliated with the Department of Electronic and Electrical Engineering, University College London, Torrington Place, WC1E 7JE, United Kingdom and he is also affiliated with Yonsei Frontier Lab, Yonsei University, Seoul, Korea (e-mail: $\rm kai\text{-}kit.wong@ucl.ac.uk$).

Ahmed Elzanaty is with the 5GIC \& 6GIC, Institute for Communication Systems (ICS), University of Surrey, GU2 7XH Guildford, U.K (e-mail:
$\rm a.elzanaty@surrey.ac.uk$).}
}

\maketitle

\begin{abstract}
With the development of the upcoming sixth-generation (6G) wireless networks, there is a pressing need for innovative technologies capable of satisfying heightened performance indicators. Fluid antenna system (FAS) is proposed recently as a promising technique to achieve higher data rates and more diversity gains by dynamically changing the positions of the antennas to form a more desirable channel. However, worries regarding the possibly harmful effects of electromagnetic (EM) radiation emitted by devices have arisen as a result of the rapid evolution of advanced techniques in wireless communication systems. Specific absorption rate (SAR) is a widely adopted metric to quantify EM radiation worldwide. In this paper, we investigate the SAR-aware multiuser multiple-input multiple-output (MIMO) communications assisted by FAS. In particular, a two-layer iterative algorithm is proposed to minimize the SAR value under signal-to-interference-plus-noise ratio (SINR) and FAS constraints. Moreover, the minimum weighted SINR maximization problem under SAR and FAS constraints is studied by finding its relationship with the SAR minimization problem. Simulation results verify that the proposed SAR-aware FAS design outperforms the adaptive backoff and fixed-position antenna designs. 
\end{abstract}


\section{Introduction}
\IEEEPARstart{A}{ wide range} of new Internet-of-Everything (IoE) applications, including industrial Internet of Things (IIoT), smart homes, autonomous vehicles, and so on, will be supported by the sixth-generation (6G) wireless networks \cite{9852292,you2021towards}. To accommodate these demanding applications efficiently, 6G networks must surpass the performance standards set by the fifth-generation (5G) networks \cite{10415645}. These standards include ultra-high data rates, low latency, extremely high reliability, as well as global coverage and connectivity. Existing 5G technologies are unable to meet these strict performance criteria, which highlights the necessity of creating novel technologies for the upcoming 6G networks \cite{9847609,9309152}.

To this end, increasing the degree of freedom (DoF) in the physical layer thus is critical. In this regard, a recently emerging technology is fluid antenna system (FAS) which provides a novel way to obtain spatial diversity \cite{wong2020fluid2,KKWong221,shojaeifard2022mimo,Wong-cl2023}. Very briefly, all types of flexible-antenna systems, both movable and non-movable, are encompassed in FAS. An up-to-date literature review can be found in \cite{Zhu-Wong-2024}. Conventional multiple-input multiple-output (MIMO) communication systems are based on fixed-position antennas (FPAs). The DoFs are thus limited by the number of antennas. In contrast, FAS-assisted MIMO systems enjoy flexibility in changing antenna positions to adaptively reconfigure the channel, thereby attaining a more favourable channel condition. This flexibility empowers FAS to have additional DoFs, hence greater spatial diversity for better performance \cite{new2023fluidw,New-twc2023}.

FAS can be implemented by using any software-controllable fluidic conductive or dielectric structure \cite{9832466,9767573,9780597}, mechanically movable antenna structure \cite{zhu2023overview}, or reconfigurable radio-frequency (RF) pixels \cite{5991914,Jing-2022} that can change its position and/or shape to reconfigure the operating frequency, radiation pattern, gain, and other features in an adaptive manner \cite{KKWong221}. The attribute relevance to this paper is the position-flexibility of FAS. Note that liquid-based movable antennas are particularly suitable for systems characterized by slow-fading channels, as the movement delay associated with these antennas makes them less effective for rapidly varying channel conditions. On the other hand, RF pixel-based non-movable antennas are more appropriate for systems operating in fast-fading channels, where the channel variations occur on much shorter time scales, requiring instantaneous adjustments. However, it is important to acknowledge that mechanically movable antennas, such as those based on microelectromechanical system (MEMS) technology, can also be employed in systems with fast-fading channels. In such systems, the movement of the antennas can be designed based on statistical channels to mitigate the effects of rapid fading \cite{10328751}. Therefore, mechanically movable antennas can be adapted to both slow- and fast-fading environments, depending on the specific design of the antenna movement and the system’s requirements.


Recent efforts have been made not only to improve channel modeling and performance analysis under different fading channels \cite{M.Khammassi23,David24, Vega-2023, Vega-2023-2, Alvim-2023, Ghadi-2023}, but also to explore flexible beamforming techniques that jointly optimize antenna positions and beamforming weights \cite{10278220,10806489,10354003}. In \cite{Psomas-dec2023}, the performance gains of a continuous FAS over the discrete counterpart were also quantified. The additional DoF using FAS has also motivated research in optimizing antenna positions for enhancing wireless communication performance. A movable antenna-assisted MIMO system was proposed in \cite{ma2023mimo}, which can be regarded as MIMO-FAS implemented by movable mechanical antenna structures. In \cite{zhu2022modeling}, a channel model was devised, and the maximum channel gain was analyzed. Multiuser cases were also recently studied by joint beamforming at the base station (BS) and FAS position optimization at the users \cite{10416896,xiao2023multiuser}. Most recently, MIMO-FAS for integrated sensing and communication (ISAC) in the downlink was tackled in \cite{Wang-fasisac2023}. Channel estimation methods for FAS were developed using compressed sensing in \cite{ma2023compressed}, sparse channel reconstruction in \cite{HXu23}, and Bayesian channel reconstruction in \cite{Dai-2023}. Statistical channel state information (CSI) has also been shown to be effective in optimizing FAS \cite{10328751}.

Moreover, developments have been made in using FAS for multiple access without the need for consecutive interference cancellation at the mobile receivers or requiring CSI at the BS side under the fluid antenna multiple access (FAMA) scheme \cite{9650760,KKWong23}. Additionally, opportunistic scheduling was considered to synergize FAMA in \cite{10078147}, showing great promise.

While future wireless networks undoubtedly need to support high data rates, often accompanied by a high density of wireless connectivity \cite{8869705}, there is growing concern that individuals will be exposed to excessive levels of electromagnetic (EM) radiation \cite{9518367,10843340}. 
The specific absorption rate (SAR), which is the absorbed power per unit mass of human tissues and is represented in watts per kilogram (${\rm W/kg}$), is a widely used metric to measure EM exposure \cite{1349969}. As a reference, the Federal Communications Commission (FCC) sets a SAR limit of $1.6~{\rm W/kg}$ for partial body exposure to safeguard human beings from elevated exposure to EM radiation \cite{8910342}. For a device with a single antenna, the SAR restriction is only an extra constraint on transmit power, which can be readily assured by lowering the transmit power below a preset threshold. However, with multiple transmit antennas, SAR is not just dependent on transmit power, and the phase differences between the antennas affect the SAR \cite{9764644,6502942}.

Several efforts have studied the SAR model to characterize the level of EM exposure. In \cite{918259}, the expression of the SAR value was proposed, which was determined by tissue density, conductivity, and electric field. In \cite{chim2004investigating}, through measurements and simulations, SAR was shown to be a sinusoidal function of the phase difference between two antennas. In \cite{6181792}, the quadratic model of SAR was proposed, and the SAR constraints were taken into consideration in the MIMO uplink system.

Optimization problems with SAR considerations exploiting the quadratic model in MIMO systems have been addressed in various scenarios \cite{9730856,ying2015closed,9444800,10167745,9576730,9774006}. In \cite{9730856}, an effective algorithm was proposed to maximize the resource efficiency under SAR constraints in the uplink MIMO system. In \cite{ying2015closed}, the authors evaluated the performance under power and SAR constraints and provided capacity analysis in multiple antenna systems. The work in \cite{9444800} then studied the design of SAR-aware precoders to maximize the energy efficiency using statistical CSI in the uplink MIMO system. In \cite{10167745}, a reconfigurable intelligent surface (RIS) was used to maximize the minimum signal-to-interference-plus-noise ratio (SINR) while adhering to the EM constraints. Additionally, the SAR was minimized in \cite{9576730} for the uplink system in a RIS-assisted environment. In \cite{9774006}, an effective method was proposed to maximize the spectral efficiency under SAR constraints in the hybrid RIS and dynamic metasurface antenna-assisted MIMO system. Most of the existing works focus on SAR reduction in uplink communication systems. However, future wireless systems operating at a higher carrier frequency, such as millimeter wave (mmWave), will entail smaller cell size and higher directional beams, which makes investigation on human exposure in downlink systems necessary. In \cite{nasim2017human}, the authors showed that SAR should be considered when estimating human RF exposure in downlink systems since 5G downlink RF fields produce significantly higher SAR. Moreover, the SAR-aware design in downlink systems was investigated in \cite{zhang2019specific}. In addition, for FAS-assisted systems, since the antenna position affects the channel, the SAR constraint is related to the antenna position, which is different from the conventional FPA-assisted systems. Therefore, the SAR-aware design in FAS-assisted downlink systems is of great importance.

Motivated by the above, we investigate the joint design of the SAR-aware precoding matrix and fluid antenna positions (a.k.a.~ports) in the FAS-assisted MIMO downlink system, which comprises of a BS with multiple fluid antennas and multiple single-FPA users. Two optimization targets are considered, namely, (1) SAR minimization and (2) SINR balancing. To the authors' best knowledge, this is the first work that investigates the SAR-aware design in FAS-assisted systems. Our main contributions are summarized as follows:
\begin{itemize}
\item First, we propose an algorithm that optimizes the antenna positions and precoding matrix to minimize the SAR value under both SINR and FAS constraints. The optimization problem is highly non-convex because of the several non-convex constraints and tightly coupled variables. Specifically, we first resort to the penalty method and divide the transformed problem into two layers. In the inner layer, the alternating optimization method is exploited to deal with the penalized problem, while in the outer layer, the penalty factor is updated.
\item Furthermore, we also investigate the minimum weighted SINR maximization under the SAR and FAS constraints in the FAS-assisted multiuser MIMO downlink system, which is closely related to the SAR minimization problem. By proving the relationship between the minimum weighted SINR maximization and the SAR minimization problems, we can handle the former by handling a series of SAR minimization problems.
\end{itemize}

The remainder of the paper is organized as follows. Section \ref{sec:model} describes the FAS-assisted multiuser MIMO downlink system model and gives the expression of SAR. Section \ref{sec:problem} formulates the SAR-aware problems, encompassing both minimum weighted SINR maximization and SAR minimization and studies the interrelationship between the two problems. Then Section \ref{sec:alg} proposes a two-layer iterative algorithm for the SAR minimization problem. Section \ref{sec:results} presents the simulation results to analyze the performance of the proposed algorithms. Finally, in Section \ref{sec:conclude}, we conclude this paper.

\textit{Notations}: $\mathbb{E}\{\cdot\}$ represents the expectation operation. The notation for the imaginary unit is $\jmath=\sqrt{-1}$. The trace is indicated by the notation $\operatorname{tr}(\cdot)$. For definitions, $\triangleq$ is the notation used. The inverse, transpose, conjugate, and conjugate-transpose of the matrix or vector are indicated by the superscripts $(\cdot)^{-1}$, $(\cdot)^{T}$, $(\cdot)^{*}$, and $(\cdot)^{H}$, respectively. The $i$-th element of vector $\ba$ is denoted by $[\mathbf{a}]_{i}$, and the $(i,j)$-th element of matrix $\mathbf{A}$ is denoted by $[\mathbf{A}]_{i, j}$. Additionally, the complex circularly symmetric Gaussian distribution with zero mean and covariance matrix $\bB$ is represented by $\mathcal{CN}(\bf 0,\bB)$. Moreover, the $l_2$ norm is represented by $\|\cdot\|_2$. The real part of the complex number is indicated by $\operatorname{Re}\{\cdot\}$. The complex number's angle and modulus are denoted by $\angle \cdot$ and $|\cdot|$.

\section{System Model}\label{sec:model}
\begin{figure}[t]
\centering
\includegraphics[scale=0.4]{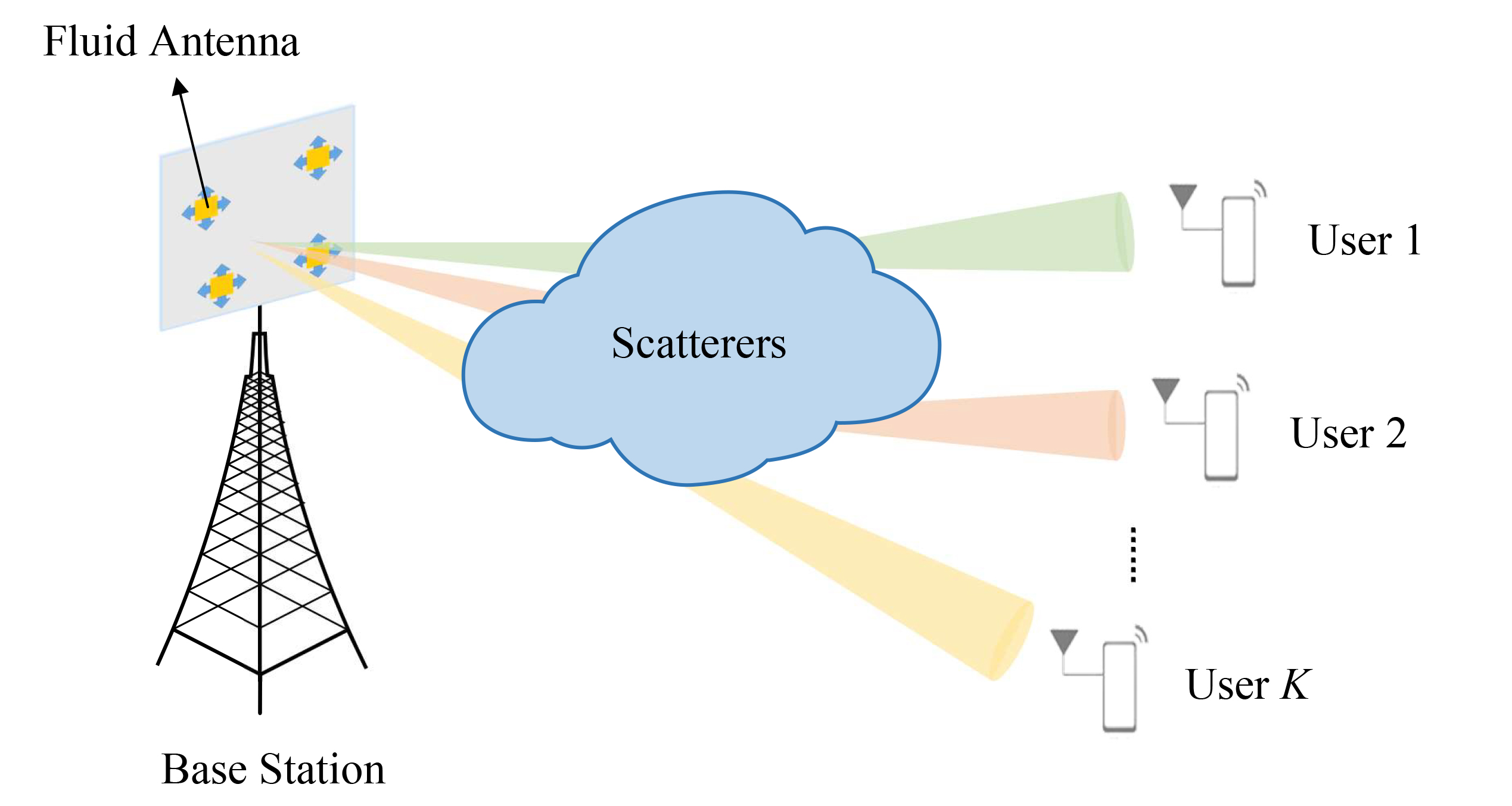}
\captionsetup{font=footnotesize}
\caption{The SAR-aware multiuser downlink MIMO system assisted by FAS.}\label{fig_model}
\end{figure}

In this section, we investigate the multiuser MIMO downlink system assisted by FAS. We focus on the narrowband communication scenario, where the symbol duration is significantly larger than the multipath delay. In the following, we illustrate the EM exposure and channel models. As shown in Fig.~\ref{fig_model}, the considered system comprises a BS with $M$ fluid antennas and $K$ single-FPA users. The fluid antennas at the BS are connected to $M$ RF chains and can switch to any positions in a given region $\mathcal {S}$. The Cartesian coordinates of the $m$-th fluid antenna are represented as $\mathbf t_m=(x_{m},y_{m})^T\in \mathcal S$ for $1\leq m\leq M$. The given region $\mathcal S$ is assumed to be a $A\times A$ square region. For $M$ transmit fluid antennas, the positions collections are represented by ${\mathbf t}=[\mathbf t_1,\dots,\mathbf t_M]\in\mathbb R^{2\times M}$.

Denote the transmitted signal from the BS as $\mathbf x=\mathbf \bP\mathbf s\in\mathbb C^{M\times 1}$, where $\mathbf s=[s_1,\dots,s_K]^T\in\mathbb C^{K\times 1}$ is the transmit data vector with covariance matrix $\mathbb{E}\{\mathbf s\mathbf s^H\}=\mathbf I_K$, and $\mathbf P=[\mathbf p_{1},\dots,\mathbf p_{K}]\in\mathbb C^{M\times K}$ is the corresponding precoding matrix. The received signal of the $k$-th user for $1\leq k\leq K$ is given by
\begin{equation}
\begin{aligned}
r_k=\mathbf h_k^H(\bt)\bP\bs+ n_k,
\end{aligned}
\end{equation}
where $\mathbf h_{k}(\bt)\in\mathbb C^{M\times 1}$ represents the channel vector between the BS with $M$ fluid antennas at positions $\bt$ and the $k$-th user with a single FPA, and $n_k\sim\mathcal{CN}(0,\sigma^2)$ is the additive white Gaussian noise.
The SINR of the $k$-th user is expressed as
\begin{equation}
\operatorname{SINR}_k(\bP,\bt)=\frac{|\mathbf h_k^H(\bt)\mathbf p_{k}|^2}{\sum_{j\neq k}|\mathbf h_k^H(\bt)\mathbf p_{j}|^2+\sigma^2}.\label{sinrk}
\end{equation}
Note that $\operatorname{SINR}_k$ in \eqref{sinrk} is a function of the transmit fluid antenna positions, $\bt$.

\subsection{EM Exposure Model}\label{ssec:exposure}
The power and EM radiation exposure level of a realistic wireless system impose limitations on the BS signal transmission. Typically, a power constraint is given by $\sum_k\mathbf p_{k}^H\mathbf p_{k} \leq P_{t}$ where $P_{t}$ represents the power budget, while for the SAR constraint, it is usually modeled as a quadratic function of the transmit signal $\bx$ to control the EM exposure at the BS. The difference between the SAR and the power constraints is that the former is dependent upon the phase changes between any two transmit antennas, whereas the latter is not \cite{chim2004investigating}. 

For a device with a single FPA, the SAR constraint is just an extra power constraint which is readily assured by lowering the power below a predetermined threshold. However, in the multi-antenna case, the SAR constraint is not just dependent on the power constraint. In this case, the SAR constraint should incorporate the SAR matrix because of the phase relationship between antennas. Here, in the FAS-assisted downlink system, we exploit the SAR constraint with a time-averaged quadratic model written as \cite{ying2015closed}
\begin{equation}
\begin{aligned}
\rm{SAR}&=\mathbb E\left\{\operatorname{tr}\left(\bx^H\bR \bx\right)\right\}\\
&=\mathbb E\left\{\operatorname{tr}\left(\sum_ks_k^H\mathbf p_{k}^H\bR \mathbf p_{k} s_k\right)\right\}\\
&=\sum_k\mathbf p_{k}^H\bR \mathbf p_{k}  \leq {Q}_0,
\end{aligned}
\end{equation}
where $\bR\in\mathbb C^{M\times M}$ is the SAR matrix of the BS to specify the dependence of SAR measurement on the transmit signal, and $Q_0$ is the SAR budget. The SAR matrix $\bR$ is a positive semidefinite matrix that characterizes the impact of the transmitted signal on electromagnetic exposure. Commonly,  the SAR
matrix is dependent on the conductivity and frequency, as well as
the employed instrument that dictates the surrounding boundary
conditions \cite{liu2013conductivity}. Unlike the conventional power constraint, the SAR constraint accounts for the phase relationships between signals transmitted from different antennas, leading to a more refined control over EM exposure. 

\begin{figure}[t]
\centering
\includegraphics[scale=0.35]{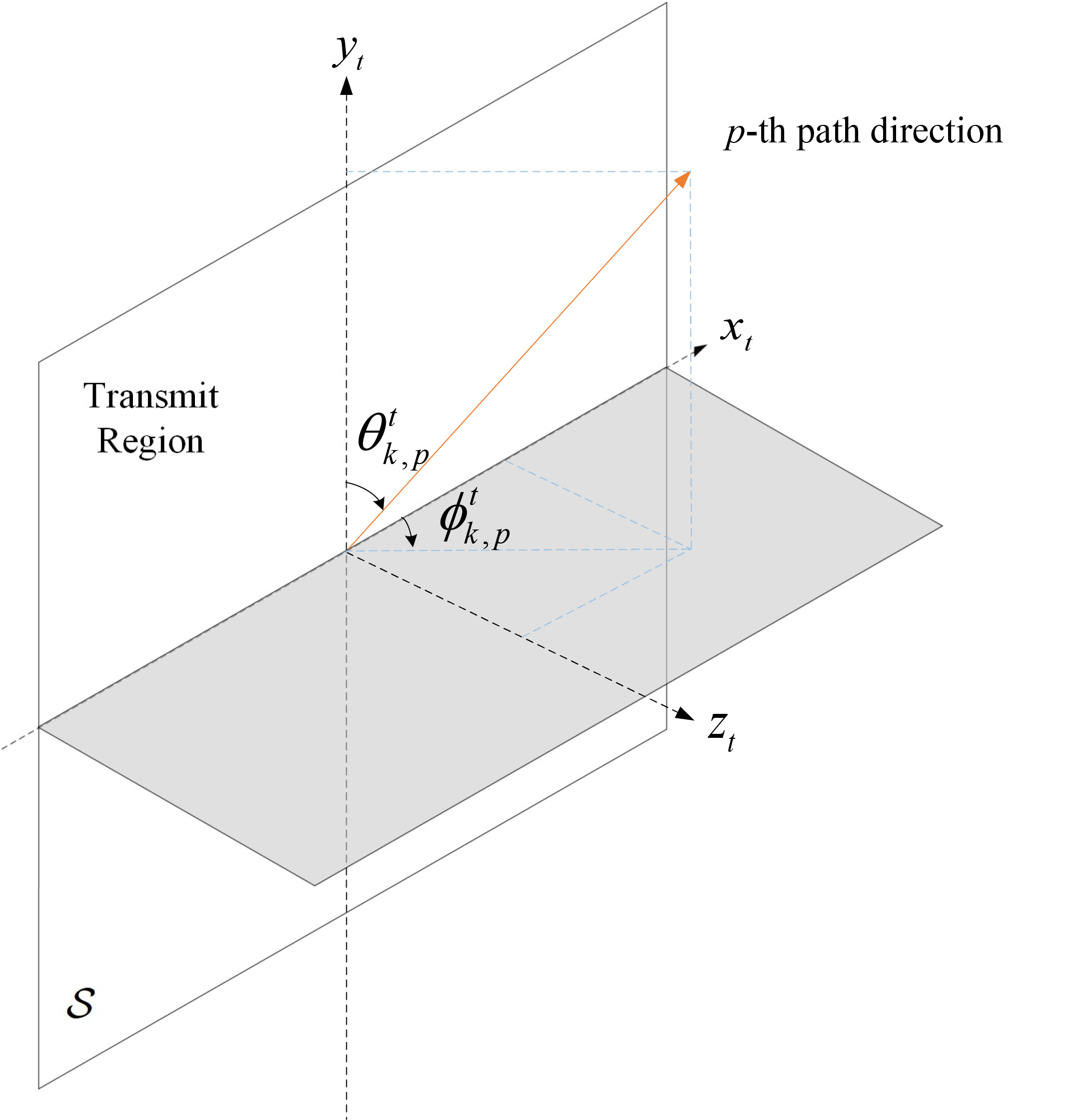}
\captionsetup{font=footnotesize}
\caption{The spatial angles for the transmit region.}\label{fig_angle}
\end{figure}

\subsection{Channel Model}\label{ssec:channel}
Denote the number of transmit paths from the BS to the $k$-th user as $L_k^t$. As seen in Fig.~\ref{fig_angle}, on the BS side, the elevation and azimuth angles of departure (AoDs) of the $p$-th transmit path from the BS to the $k$-th user are $\theta_{k,p}^t$ and $\phi_{k,p}^t$, respectively, for $1\leq p\leq L_k^t$. The propagation distance difference in the $p$-th transmit path for the $k$-th user between the origin and the position of the $m$-th transmit fluid antenna is \cite{ma2023mimo}
\begin{equation}
\rho_{k,p}^t(\mathbf{t}_m)=x_m \sin \theta_{k,p}^t \cos \phi_{k,p}^t+y_m \cos \theta_{k,p}^t.
\end{equation}
The corresponding transmit field response vector is given by \cite{ma2023mimo}
\begin{equation}
\mathbf{g}_k(\mathbf t) \triangleq\left[e^{\jmath\frac{2\pi}{\lambda}\rho_{k,1}^t(\mathbf{t})},\dots,e^{\jmath\frac{2\pi}{\lambda}\rho_{k,L_t}^t(\mathbf{t})}\right]^T \in \mathbb{C}^{L_k^t\times 1}.
\end{equation}
where $\lambda$ is the signal wavelength. The field response matrix of $M$ transmit fluid antennas is written as
\begin{equation}
\mathbf{G}_k(\bt) \triangleq\left[\mathbf{g}_k\left(\mathbf{t}_1\right), \mathbf{g}_k\left(\mathbf{t}_2\right), \ldots, \mathbf{g}_k\left(\mathbf{t}_M\right)\right] \in \mathbb{C}^{L_k^t \times M}.
\end{equation}
In addition, the path response vector is denoted as $\mathbf{f}_k=[f_{k,1},f_{k,2},\dots,f_{k,L_k^t}]^H\in \mathbb{C}^{L_k^t \times 1}$, which represents the multi-path response coefficients from the BS to the $k$-th user. The channel vector $\mathbf h_k({\mathbf t})$ from the BS with fluid antennas at positions $\bt$ to the $k$-th user with a single FPA is written as \cite{ma2023mimo,xiao2023multiuser}
\begin{equation}
\mathbf h_k({\mathbf t})=\mathbf{G}_k(\bt)^H\mathbf{f}_k .
\end{equation}

\section{Problem Formulation}\label{sec:problem}
In this section, we aim to maximize the minimum weighted SINR by optimizing the precoding matrix $\bP$ and the positions of the transmit fluid antennas $\bt$. To limit the level of EM exposure, we impose the SAR constraint. Moreover, to prevent the mutual coupling effect, a minimum distance $D$ between the transmit fluid antennas is introduced. The SINR balancing problem is formulated as
\begin{equation}
\begin{aligned}
\mathcal{P}_1:\max _{\bP,\bt} 
&\quad\min_{k} \frac{1}{\gamma_k}\operatorname{SINR}_k(\bP,\bt) \\
\text { s. t. } 
& \quad\sum_k\mathbf p_{k}^H\bR \mathbf p_{k}  \leq {Q}_0,\\
& \quad{\mathbf t} \in \mathcal{S},\\
&\quad\left\|\mathbf{t}_m-\mathbf{t}_l\right\|_2 \geq D,\quad m,l=1,\dots,M,\quad m\neq l,
\end{aligned}
\end{equation}
where $\gamma_k$ is the weight parameter for the $k$-th user. To find a more tractable formulation, we first introduce an auxiliary variable $\beta_0$ and reformulate the problem $\mathcal{P}_1$ as
\begin{equation}
\begin{aligned}
\mathcal{P}_2:\max _{\bP,\bt,\beta_0} 
&\quad \beta_0 \\
\text { s. t. } 
& \quad \frac{|\mathbf h_k^H(\bt)\mathbf p_{k}|^2}{\sum_{j\neq k}|\mathbf h_k^H(\bt)\mathbf p_{j}|^2+\sigma^2}\geq \beta_0\gamma_k, \quad\forall k,\\
& \quad\sum_k\mathbf p_{k}^H\bR \mathbf p_{k}  \leq {Q}_0,\\
& \quad{\mathbf t} \in \mathcal{S},\\
&\quad\left\|\mathbf{t}_m-\mathbf{t}_l\right\|_2 \geq D,\quad m,l=1,\dots,M,\quad m\neq l.
\end{aligned}
\end{equation}

Note that the optimal precoding matrix $\bP$ and the fluid antenna positions $\bt$ of $\mathcal{P}_2$ are equivalent to those of $\mathcal{P}_1$ since the first inequality constraint of $\mathcal{P}_2$ holds with equality at the optimum.

Alternatively, we also consider the SAR minimization problem, which is a closely related problem of $\mathcal{P}_1$. In this problem, we aim to minimize the SAR value under the FAS and SINR constraints, which is given by
\begin{equation}
\begin{aligned}
\mathcal{P}_{3}: \min _{\bP,\bt} 
&\quad\sum_k\mathbf p_{k}^H\bR \mathbf p_{k} \\
\text { s. t.} 
& \quad\frac{|\mathbf h_k^H(\bt)\mathbf p_{k}|^2}{\sum_{j\neq k}|\mathbf h_k^H(\bt)\mathbf p_{j}|^2+\sigma^2}\geq \bar{\gamma}_k=\beta_0\gamma_k, \quad\forall k,\\
& \quad{\mathbf t} \in \mathcal{S},\\
&\quad\left\|\mathbf{t}_m-\mathbf{t}_l\right\|_2 \geq D,\quad m,l=1,\dots,M,\quad m\neq l,
\end{aligned}
\end{equation}
where $\beta_0$ is the SINR threshold. 

Denote the optimal objective function value of $\mathcal{P}_{2}$ with given $Q_0$ as $\mathcal{P}_{2}\left(Q_0\right)$ and the optimal objective function value of $\mathcal{P}_{3}$ with given $\beta_0$ as $\mathcal{P}_{3}\left(\beta_0\right)$, respectively. In the following lemma, we show the relationship between problem $\mathcal{P}_2$ and problem $\mathcal{P}_3$.

\begin{algorithm}[t]
\caption{Bisection Search Method to Handle Problem $\mathcal{P}_{2}$}
\label{a_1}
 \begin{algorithmic}[1]
 \State Initialize the accuracy $\varepsilon_1$, the lower bound $\beta_{l}$ and upper bound $\beta_{u}$.
\Repeat
\State Calculate $\beta_0=(\beta_{l}+\beta_{u})/2$.
\State Tackle problem $\mathcal{P}_{3}$.
\If {$\sum_k\mathbf p_{k}^H\bR \mathbf p_{k} \leq Q_0$}
\State Set $\beta_{l}=\beta_0$. 
\Else 
\State Set $\beta_{u}=\beta_0$.
\EndIf
\Until {$|\beta_{l}-\beta_{u}|\leqslant \varepsilon_1$.}
 \end{algorithmic}
\end{algorithm}

\emph{Lemma 1:} The minimum weighted SINR maximization problem $\mathcal{P}_2$ and the SAR minimization problem $\mathcal{P}_3$ have the following relationship:
\begin{align}
\beta_0 & =\mathcal{P}_2\left(\mathcal{P}_{3}\left(\beta_0\right)\right), \label{inverse1}\\
Q_0& =\mathcal{P}_{3}\left(\mathcal{P}_2\left(Q_0\right)\right).\label{inverse2}
\end{align}

\emph{Proof:} Following the approach in \cite{wiesel2005linear}, we begin by proving \eqref{inverse1} by contradiction. Assume the contrary, i.e., $Q$, $\mathbf{P}$, $\bt$ are the optimal value and argument of $\mathcal{P}_{3}(\beta)$, and $\tilde{\beta} \neq \beta$, $\tilde{\mathbf{P}}$, $\tilde{\mathbf{t}}$ are the optimal value and argument of $\mathcal{P}_{2}(Q)$. If $\tilde{\beta}<\beta$, there is a contradiction for the optimality of $\tilde{\mathbf{P}}$ and $\tilde{\mathbf{t}}$ for $\mathcal{P}_{2}(Q)$, since $\mathbf{P}$ and $\mathbf{t}$ are feasible and they provide a larger objective value $\beta$. Otherwise, if $\tilde{\beta}>\beta$, there is a contradiction for the optimality of $\mathbf{P}$ for $\mathcal{P}_{3}(\beta)$, since $\tilde{\beta}>\beta$, and we can always find $c<1$ such that $c \tilde{\mathbf{P}}$ will still be feasible but will result in a smaller objective. The proof of \eqref{inverse2} is similar and is therefore omitted. Thus, the proof is complete.\hfill $\blacksquare$

From \emph{Lemma 1}, if $Q_0=\mathcal{P}_{3}(\beta_0)$, the solutions of $\mathcal{P}_{3}(\beta_0)$ will also be optimal for $\mathcal{P}_{2}(Q_0)$ \cite{wiesel2005linear}. By exploiting \emph{Lemma 1} and the fact that $\mathcal{P}_2$ is quasiconvex over $\beta_0$, a bisection search over $\beta_0$ can be used to handle $\mathcal{P}_2$ after $\mathcal{P}_3$ is tackled \cite{zhang2019specific}, which is detailed in \textbf{Algorithm \ref{a_1}}. Therefore, in the following, we focus on handling problem $\mathcal{P}_{3}$.

The challenges in tackling $\mathcal{P}_{3}$ come from the coupled variables and non-convex constraints. In the sequel, we propose an effective method to decouple $\mathcal{P}_{3}$ into several sub-problems.

\section{Two-Layer Iterative Algorithm}\label{sec:alg}
\begin{figure}[t]
\centering
\includegraphics[scale=0.53]{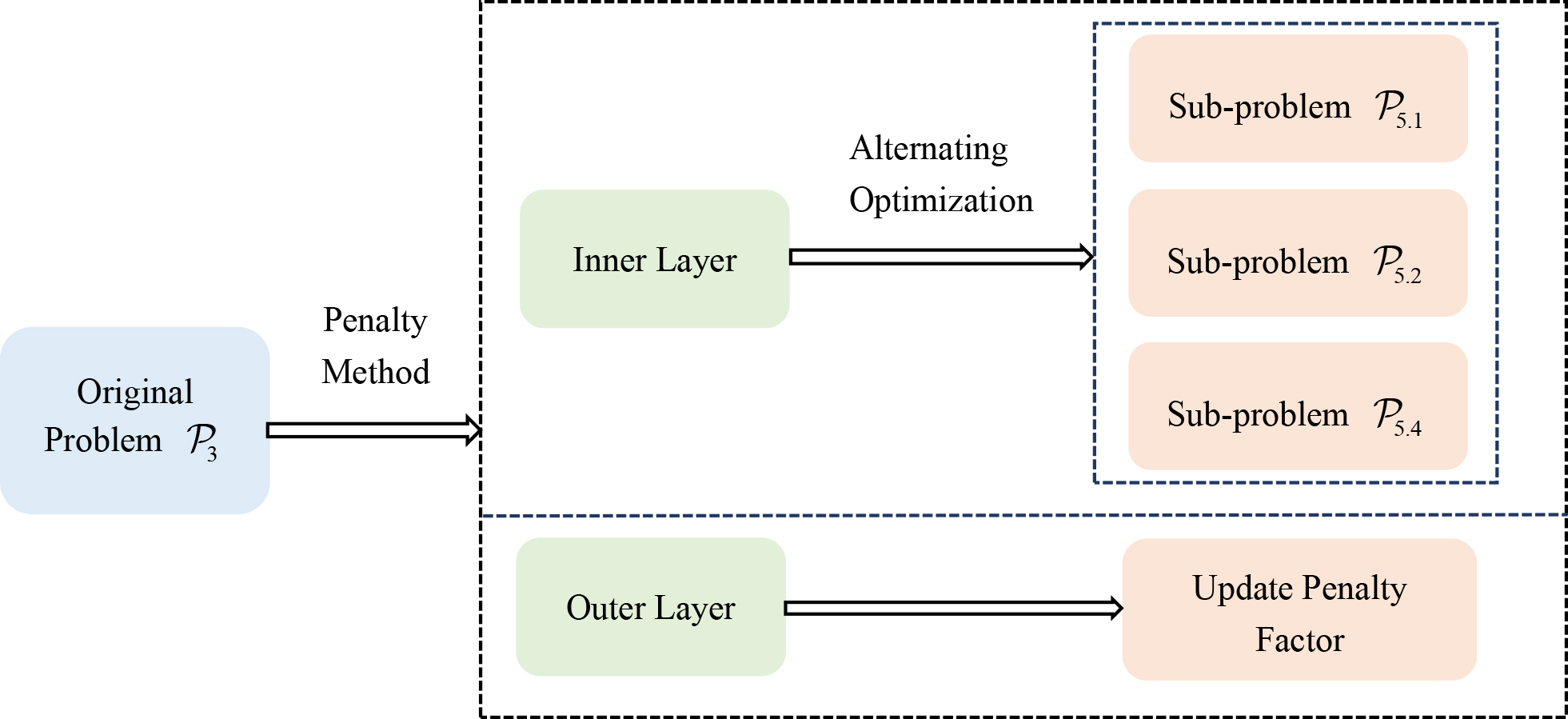}
\captionsetup{font=footnotesize}
\caption{Illuastration of the two-layer iterative algorithm.}\label{fig_algorithm}
\end{figure}

Here, a two-layer iterative algorithm is proposed to tackle the SAR minimization problem $\mathcal{P}_{3}$. The penalty method is first employed to decouple the precoding matrix $\bP$ and the transmit fluid antenna positions $\bt$ in the SINR constraints. The original SAR minimization problem becomes a two-layer problem where the penalized problem in the inner layer can be handled by the alternating optimization method, and the penalty factor in the outer layer is updated. The overall algorithm for handling problem $\mathcal P_3$ is summarized in Fig.~\ref{fig_algorithm}, which will be detailed below.

\subsection{Problem Transformation}\label{ssec:transform}
To handle the SAR minimization problem $\mathcal{P}_{3}$, we first deal with the non-convex SINR constraints. To decouple the tightly coupled optimizing variables in the constraints, new auxiliary variables $z_{k,j}$ are introduced to represent $\mathbf h_{k}^H(\bt)\mathbf p_{j}$. The SINR constraints can be reformulated as
\begin{align}
\frac{|z_{k,k}|^2}{\sum_{j\neq k}|z_{k,j}|^2+\sigma^2}&\geq \bar{\gamma}_k,~\forall k,\label{sinr1}\\
z_{k,j}&=\mathbf h_{k}^H(\bt)\mathbf p_{j},~\forall k,j.\label{sinr2}
\end{align}
By replacing the original SINR constraints with \eqref{sinr1} and \eqref{sinr2}, the original problem $\mathcal P_3$ is reformulated as
\begin{equation}
\begin{aligned}
\mathcal{P}_{4}: \min _{\bP, \bt,z_{k,j}}
&\quad\sum_k\mathbf p_{k}^H\bR \mathbf p_{k} \\
\text { s. t. }
& \quad\frac{|z_{k,k}|^2}{\sum_{j\neq k}|z_{k,j}|^2+\sigma^2}\geq \bar{\gamma}_k, \quad \forall k,\\
& \quad z_{k,j}=\mathbf h_{k}^H(\bt)\mathbf p_{j},\quad \forall k,j,\\
& \quad{\mathbf t} \in \mathcal{S},\\
&\quad\left\|\mathbf{t}_m-\mathbf{t}_l\right\|_2 \geq D,\quad m,l=1,\dots,M,\quad m\neq l.
\end{aligned}
\end{equation}
Then, we exploit the penalty approach to deal with the equality constraints. The transformed problem $\mathcal{P}_{4}$ can be rewritten as
\begin{equation}
\begin{aligned}
\mathcal{P}_{5}: \min _{\bP,\bt,z_{k,j}}
&\quad\sum_k\mathbf p_{k}^H\bR \mathbf p_{k}+\mu\sum_{k=1}^K\sum_{j=1}^K|\mathbf h_{k}^H(\bt)\mathbf p_{j}- z_{k,j}|^2 \\
\text { s. t. }
& \quad\frac{|z_{k,k}|^2}{\sum_{j\neq k}|z_{k,j}|^2+\sigma^2}\geq \bar{\gamma}_k, \quad k=1, \ldots, K,\\
& \quad{\mathbf t} \in \mathcal{S},\\
&\quad\left\|\mathbf{t}_m-\mathbf{t}_l\right\|_2 \geq D,\quad m,l=1,\dots,M,\quad m\neq l,
\end{aligned}
\end{equation}
where $\mu$ is the penalty factor. Theoretically, the penalty factor $\mu$ needs to be large so that the equality constraints hold. But the penalty term dominates the objective function when $\mu$ is very large, which means the optimization problem is not to minimize the SAR value. Therefore, the penalty factor $\mu$ should be initialized to a very small value and gradually increase \cite{9115725}. In so doing, a high-precision solution satisfying the equality constraints can be obtained.

After utilizing the penalty method, the problem $\mathcal{P}_{5}$ is still hard to handle because of the non-convex constraints. We then propose a two-layer iterative algorithm to deal with the inner and outer layers. The penalized problem is tackled through the alternating optimization method in the inner layer, while the penalty factor is updated in the outer layer.

\subsection{Inner Layer}\label{ssec:inner}
\subsubsection{Optimize $\bP$}With given $\{z_{k,j}\}_{k,j=1}^K$ and $\bt$, the problem $\mathcal{P}_{5}$ is rewritten as
\begin{equation}
\mathcal{P}_{5.1}:  \min _{\bP} 
\sum_k\mathbf p_{k}^H\bR \mathbf p_{k}+\mu\sum_{k=1}^K\sum_{j=1}^K|\mathbf h_{k}^H(\bt)\mathbf \bp_j- z_{k,j}|^2.
\end{equation}
Since $\mathcal{P}_{5.1}$ is an unconstrained convex optimization problem, by setting the first-order derivative of the objective function regarding the variable $\bp_k$ to zero matrix, the optimal solution can be obtained, given by
 \begin{equation}
\bp_k=\left(\bR+\bR^H+2\mu\sum_{i=1}^K\mathbf h_{i}(\bt)\mathbf h_{i}^H(\bt)\right)^{-1}\ba_k^H,
\end{equation}
where
\begin{equation}
\ba_k=2\mu\sum_{i=1}^K z_{i,k}^{*}\mathbf h_{i}^H(\bt).
\end{equation}

\subsubsection{Optimize $\{z_{k,j}\}_{k,j=1}^K$}For given $\bP$ and $\bt$, $\{z_{k,j}\}_{k,j=1}^K$ in problem $\mathcal P_5$ can be optimized by handling
\begin{equation}
\begin{aligned}
\mathcal{P}_{5.2}:  \min _{\{z_{k,j}\}_{k,j=1}^K} 
&\quad\sum_{k=1}^K\sum_{j=1}^K|\mathbf h_{k}^H(\bt)\mathbf p_{j}- z_{k,j}|^2 \\
\text { s. t. } 
& \quad\frac{|z_{k,k}|^2}{\sum_{j\neq k}|z_{k,j}|^2+\sigma^2}\geq \bar{\gamma}_k, \quad k=1, \ldots, K.
\end{aligned}
\end{equation}
Note that the optimization variables in relation to different users $z_{k,j}$, $k=1,2,\dots,K$ can be directly separated in the objective function and SINR constraints. Therefore, by dealing with $K$ independent sub-problems, each with a single SINR constraint, we can handle $\mathcal{P}_{5.2}$. The sub-problem related to the $k$-th user is expressed as
\begin{equation}
\begin{aligned}
\mathcal{P}_{5.3}:  \min _{\{z_{k,j}\}_{j=1}^K} 
&\quad\sum_{j=1}^K|\mathbf h_{k}^H(\bt)\mathbf p_{j}- z_{k,j}|^2 \\
\text { s. t. } 
& \quad\frac{|z_{k,k}|^2}{\sum_{j\neq k}|z_{k,j}|^2+\sigma^2}\geq \bar{\gamma}_k.
\end{aligned}
\end{equation}
Although problem $\mathcal{P}_{5.3}$ is a non-convex problem, the strong duality is demonstrated to hold since it satisfies the Slater's condition \cite{boyd2004convex,9133435}. Then, we resort to the Lagrange duality method. The Lagrangian function of problem ${P}_{5.3}$ can be expressed as follows by denoting the Lagrange multiplier related to the $k$-th SINR constraint as $\zeta_k$. 
\begin{multline}
\mathcal{L}\left(\{z_{k,j}\}_{j=1}^K, \zeta_k\right)=\left(1-\zeta_k\right)\left|z_{k, k}\right|^2
+\sum_{j=1\atop j \neq k}^K\left(1+\zeta_k \bar{\gamma}_k\right)\left|z_{k, j}\right|^2 \\
 -2 \sum_{j=1}^K \operatorname{Re}\left\{\mathbf h_{k}^H(\bt)\mathbf p_{j} z_{k, j}^*\right\}.
\end{multline}
The optimal solution can be obtained by using the first-order optimality condition, which is given by
\begin{equation}
\begin{aligned}
z_{k, k}(\zeta_k) & =\frac{\mathbf h_{k}^H(\bt)\mathbf p_{k}}{1-\zeta_k}, \\
z_{k, j}(\zeta_k) & =\frac{\mathbf h_{k}^H(\bt)\mathbf p_{j}}{1+\zeta_k \bar{\gamma}_k}, \quad 1\leq j\neq k\leq K.\label{dual_xki}
\end{aligned}
\end{equation}
The Lagrange multiplier $\zeta_k$ should satisfy the complementary slackness condition for the SINR constraint, which can be simplified as
\begin{equation}
\begin{aligned}
\zeta_k\left(\frac{|\mathbf h_{k}^H(\bt)\mathbf p_{k}|^2}{(1-\zeta_k)^2}-\sum_{j\neq k}\frac{\bar{\gamma}_k|\mathbf h_{k}^H(\bt)\mathbf p_{j}|^2}{(1+\zeta_k \bar{\gamma}_k)^2}
-\bar{\gamma}_k\sigma^2\right)=0.
\end{aligned}
\end{equation}
If $z_{k, k}(0)$ and $z_{k, j}(0),\forall j \neq k$ satisfy the SINR constraint, then $\zeta_k=0$. Otherwise, we need to find $\zeta_k$ that satisfies the following equality
\begin{equation}
\begin{aligned}
Y(\zeta_k)=&\frac{|\mathbf h_{k}^H(\bt)\mathbf p_{k}|^2}{(1-\zeta_k)^2}-\sum_{j\neq k}\frac{\bar{\gamma}_k|\mathbf h_{k}^H(\bt)\mathbf p_{j}|^2}{(1+\zeta_k \bar{\gamma}_k)^2}
-\bar{\gamma}_k\sigma^2=0.
\end{aligned}
\end{equation}
Since $Y(\zeta_k)$ with respect to $\zeta_k$ is a monotonically increasing function for $\zeta_k\leq 1$, the optimal $\zeta_k$ can be obtained through the bisection search method \cite{9133435}.

\subsubsection{Optimize $\bt$}With fixed $\bP$ and $\{z_{k,j}\}_{k,j=1}^K$, the corresponding transmit fluid antenna positions $\bt$ optimization problem can be expressed as
\begin{equation}
\begin{aligned}
\mathcal{P}_{5.4}:  \min _{\bt} 
&\quad\sum_{k=1}^K\sum_{j=1}^K|\mathbf h_{k}^H(\bt)\mathbf p_{j}- z_{k,j}|^2 \\
\text { s. t. } 
& \quad{\mathbf t} \in \mathcal{S},\\
&\quad\left\|\mathbf{t}_m-\mathbf{t}_l\right\|_2 \geq D,\quad m,l=1,\dots,M,\quad m\neq l.
\end{aligned}
\end{equation}
By expanding the objective function, problem $\mathcal{P}_{5.4}$ is reformulated as
\begin{equation}
\begin{aligned}
\mathcal{P}_{5.5}:  \min _{\bt} 
&\quad\sum_{k=1}^K\sum_{j=1}^K\mathbf h_{k}^H(\bt)\mathbf p_{j}\mathbf p_{j}^H\mathbf h_{k}(\bt)-\\&\quad\quad\quad\quad 2 \operatorname{Re}\left\{\mathbf h_{k}^H(\bt)\mathbf p_{j} z_{k,j}^*\right\}+|z_{k,j}|^2 \\
\text { s. t. } 
& \quad{\mathbf t} \in \mathcal{S},\\
&\quad\left\|\mathbf{t}_m-\mathbf{t}_l\right\|_2 \geq D,\quad m,l=1,\dots,M,\quad m\neq l.
\end{aligned}
\end{equation}

\begin{figure*}[hb]
\centering\hrulefill
\begin{multline}\label{barq}
\bar{q}\left(\bt_m\right)=\sum_{k=1}^K\sum_{j=1}^K\left[\sum_{l\neq m}^M[\bW_{j}]_{m,l}\mathbf{f}_k^H\bg_k(\bt_m)\bg_k^H(\bt_l)\mathbf{f}_k+\left(\sum_{l\neq m}^M[\bW_{j}]_{l,m}\mathbf{f}_k^H\bg_k(\bt_{l})+[\bW_{j}]_{m,m}\mathbf{f}_k^H\bg_k(\bt_{m})\right)\bg_k^H(\bt_m)\mathbf{f}_k\right.\\
\left.-\sum_{p=1}^{L_k^t}2|[\mathbf{f}_k^H]_{p}||[\bar{\bp}_{k,j}]_m|\cos(J_3(p,j,m,k))\right]
\end{multline}
\end{figure*}

Since the $M$ fluid antenna positions $\{\bt_m\}_{m=1}^M$ are coupled in the objective function, we alternately optimize one variable $\bt_m$ while keeping the others $\{\bt_l\}_{l\neq m, l=1}^M$ fixed. With given $\{\bt_l\}_{l\neq m, l=1}^M$, minimizing the objective function in $\mathcal{P}_{5.5}$ is equivalent to minimizing $\bar{q}\left(\bt_m\right)$ in \eqref{barq} (see bottom of this page) where $\bW_j=\bp_j\bp_j^H$. Since $\bar{q}\left(\bt_m\right)$ is a non-convex function regarding $\bt_m$, the successive convex approximation (SCA) method is exploited to derive an upper bound of $\bar{q}\left(\bt_m\right)$, which is given by \cite{10416896,magnus2019matrix}
\begin{equation}
\begin{aligned}
\bar{q}\left(\bt_m\right) &\leq \bar{q}\left(\bt_m^{(n)}\right)+\nabla \bar{q}\left(\bt_m^{(n)}\right)^T\left(\bt_m-\bt_m^{(n)}\right)+\\
&\quad\frac{\tau_m}{2}\left(\bt_m-\bt_m^{(n)}\right)^T\left(\bt_m-\bt_m^{(n)}\right) \\
&=\frac{\tau_m}{2} \bt_m^T \bt_m+\left(\nabla \bar{q}\left(\bt_m^{(n)}\right)-\tau_m \bt_m^{(n)}\right)^T \bt_m+\\
&\quad\underbrace{\bar{q}\left(\bt_m^{(n)}\right)+\frac{\tau_m}{2}\left(\bt_m^{(n)}\right)^T \bt_m^{(n)}}_{\text {constant }},\label{secondtaylor}
\end{aligned}
\end{equation}
where $n$ is the iteration index, $\nabla \bar{q}\left(\bt_m\right)$ is the gradient vector of $\bar{q}\left(\bt_m\right)$ over $\bt_m$, $\tau_m$ is a positive real number that satisfies $\tau_m\bI_2\succeq \nabla^2 \bar{q}\left(\bt_m\right)$, and $\nabla^2 \bar{q}\left(\bt_m\right)$ denotes the Hessian matrix of $\bar{q}\left(\bt_m\right)$ over $\bt_m$. The elements of $\nabla \bar{q}\left(\bt_m\right)$ and $\nabla^2 \bar{q}\left(\bt_m\right)$ are given in \eqref{gra1}--\eqref{Hes3} on the next two pages, where 
\begin{figure*}[hb]
\centering
\hrulefill
\begin{multline}\label{gra1}
 \frac{\partial \bar{q}\left(\boldsymbol{t}_m\right)}{\partial x_m}=\frac{2\pi}{\lambda}\sum_{k=1}^K\sum_{j=1}^K\left[-2\sum_{p=1}^{L_k^t}\sum_{p_1=1}^{L_k^t}\sum_{l\neq m}^MT(p,p_1,k)|[\bW_{j}]_{m,l}|\sin\left(J_1(k,p,l,p_1,j,m)\right)\sin\theta_{k,p_1}^t\cos\phi_{k,p_1}^t\right.\\
\left.-[\bW_{j}]_{m,m}\sum_{p_2=1}^{L_k^t}\sum_{p_1=1}^{L_k^t}T(p_2,p_1,k)\sin\left(J_2(k,p_1,p_2,m)\right)+2\sum_{p=1}^{L_k^t}|[\mathbf{f}_k^H]_{p}||[\bar{\bp}_{k,j}]_m|\sin\left(J_3(p,j,m,k)\right)\sin\theta_{k,p}^t\cos\phi_{k,p}^t\right]
\end{multline}
\begin{multline}\label{gra2}
\frac{\partial \bar{q}\left(\boldsymbol{t}_m\right)}{\partial y_m}=\frac{2\pi}{\lambda}\sum\limits_{k=1}^K\sum_{j=1}^K\left[-2\sum\limits_{l\neq m}^M\sum\limits_{p=1}^{L_k^t}\sum\limits_{p_1=1}^{L_k^t}T(p,p_1,k)|[\bW_{j}]_{m,l}|\sin\left(J_1(k,p,l,p_1,j,m)\right)\cos\theta_{k,p_1}^t\right.\\
-[\bW_{j}]_{m,m}\sum\limits_{p_2=1}^{L_k^t}\sum\limits_{p_1=1}^{L_k^t}T(p_2,p_1,k)\sin\left(J_2(k,p_1,p_2,m)\right)\left(\cos\theta_{k,p_1}^t-\cos\theta_{k,p_2}^t\right)\\
\left.+2\sum\limits_{p=1}^{L_k^t}|[\mathbf{f}_k^H]_{p}||[\bar{\bp}_{k,j}]_m|\sin(J_3(p,j,m,k))\cos\theta_{k,p}^t\right]
\end{multline}
\end{figure*}
\begin{figure*}[hb]
\centering\hrulefill
\begin{multline}\label{Hes1}
 \frac{\partial^2 \bar{q}\left(\boldsymbol{t}_m\right)}{\partial x_m^2}=\frac{4\pi^2}{\lambda^2}\sum\limits_{k=1}^K\sum_{j=1}^K\left[-2\sum\limits_{l\neq m}^M\sum\limits_{p_1=1}^{L_k^t}\sum\limits_{p=1}^{L_k^t}T(p,p_1,k)|[\bW_{j}]_{m,l}|\cos\left(J_1(k,p,l,p_1,j,m)\right)\sin^2\theta_{k,p_1}^t\cos^2\phi_{k,p_1}^t\right.\\
 -[\bW_{j}]_{m,m}\sum\limits_{p_1=1}^{L_k^t}\sum\limits_{p_2=1}^{L_k^t}T(p_2,p_1,k)\cos\left(J_2(k,p_1,p_2,m)\right) \left(\sin\theta_{k,p_1}^t\cos\phi_{k,p_1}^t-\sin\theta_{k,p_2}^t\cos\phi_{k,p_2}^t\right)^2\\
 \left.+2\sum\limits_{p=1}^{L_k^t}|[\mathbf{f}_k^H]_{p}||[\bar{\bp}_{k,j}]_m|\cos(J_3(p,j,m,k))\sin^2\theta_{k,p}^t\cos^2\phi_{k,p}^t\right]
\end{multline}
\begin{multline}\label{Hes2}
\frac{\partial^2 \bar{q}\left(\boldsymbol{t}_m\right)}{\partial y_m^2}=\frac{4\pi^2}{\lambda^2}\sum\limits_{k=1}^K\sum_{j=1}^K\left[-2\sum\limits_{l\neq m}^M\sum\limits_{p_1=1}^{L_k^t}\sum\limits_{p=1}^{L_k^t}T(p,p_1,k)|[\bW_{j}]_{m,l}|\cos\left(J_1(k,p,l,p_1,j,m)\right)\cos^2\theta_{k,p_1}^t\right.\\
-[\bW_{j}]_{m,m}\sum\limits_{p_1=1}^{L_k^t}\sum\limits_{p_2=1}^{L_k^t}T(p_2,p_1,k)\cos\left(J_2(k,p_1,p_2,m)\right)\left(\cos\theta_{k,p_1}^t-\cos\theta_{k,p_2}^t\right)^2\\
\left.+2\sum\limits_{p=1}^{L_k^t}|[\mathbf{f}_k^H]_{p}||[\bar{\bp}_{k,j}]_m|\cos(J_3(p,j,m,k))\cos^2\theta_{k,p}^t\right]
\end{multline}
\begin{multline}\label{Hes3}
\frac{\partial^2 \bar{q}\left(\boldsymbol{t}_m\right)}{\partial x_m\partial y_m}=\frac{4\pi^2}{\lambda^2}\sum_{k=1}^K\sum_{j=1}^K\left[-2\sum_{l\neq m}^M\sum_{p_1=1}^{L_k^t}\sum_{p=1}^{L_k^t}T(p,l_2,k)|[\bW_{j}]_{m,l}|\cos\left(J_1(k,p,l,p_1,j,m)\right)\sin\theta_{k,p_1}^t\cos\phi_{k,p}^t\cos\theta_{k,p_1}^t\right.\\
-[\bW_{j}]_{m,m}\sum\limits_{p_1=1}^{L_k^t}\sum\limits_{p_2=1}^{L_k^t}T(p_2,p_1,k)\cos\left(J_2(k,p_1,p_2,m)\right)\\
\times\left[\sin\theta_{k,p_1}^t\cos\phi_{k,p_1}^t\left(\cos\theta_{k,p_1}^t-\cos\theta_{k,p_2}^t\right)-\sin\theta_{k,p_2}^t\cos\phi_{k,p_2}^t(\cos\theta_{k,p_1}^t-\cos\theta_{k,p_2}^t)\right]\\
\left.+2\sum\limits_{p=1}^{L_k^t}|[\mathbf{f}_k^H]_{p}||[\bar{\bp}_{k,j}]_m|\cos(J_3(p,j,m,k))\sin\theta_{k,p}^t\cos\phi_{k,p}^t\cos\theta_{k,p}^t\right]
\end{multline}
\end{figure*}
\begin{align}
\bar{\bp}_{k,j}&=\bp_j z_{k,j}^{*},\\
J_1(k,l,l_1,l_2,j,m)&=\angle[\mathbf{f}_k]_{l}+\angle[\mathbf{f}_k^H]_{l_2}+\angle[\bW_{j}]_{m ,l_1}\nonumber\\
&+\frac{2\pi}{\lambda}\left(\rho_{k,l_2}^t(\bt_m)-\rho_{k,l}^t(\bt_{l_1})\right),\\
J_2(k,l_1,l_2,m)&=\angle[\mathbf{f}_k^H]_{l_1}+\angle[\mathbf{f}_k]_{l_2}\nonumber\\
&+\frac{2\pi}{\lambda}\left(\rho_{k,l_1}^t(\bt_m)-\rho_{k,l_2}^t(\bt_{m})\right),\\
J_3(l,j,m,k)&=\angle[\mathbf{f}_k^H]_{l}+\angle[\bar{\bp}_{k,j}]_m+\frac{2\pi}{\lambda}\rho_{k,l}^t(\bt_m),\\
T(l,l_2,k)&=|[\mathbf{f}_k]_{l}||[\mathbf{f}_k^H]_{l_2}|.
\end{align}

\begin{algorithm}[t]
\caption{SCA-Based Method to Handle Problem $\mathcal{P}_{5.4}$}
\label{a_3}
 \begin{algorithmic}[1]
 \State Initialize the thresholds $\varepsilon_2$, the iteraton index $n=0$ and fluid antenna positions $\bt$.
\For {m = 1: M}
\Repeat 
\State Obtain $\nabla \bar{q}\left(\bt_m^{(n)}\right)$ according to \eqref{gra1} and \eqref{gra2}.
\State Obtain $\tau_m$ as \eqref{deltam}.
\State Update $\bt_{m}^{(n+1)}$ as \eqref{tmn1}.
\If{$\bt_{m}^{(n+1)}$ does not satisfy \eqref{cons1} or \eqref{cons2}}
\State Update $\bt_{m}^{(n+1)}$ by handling problem $\mathcal{P}_{5.7}$.
\EndIf
\State $n=n+1$.
 \Until{The decrease of the objective value of $\mathcal{P}_{5.4}$ is below a threshold $\varepsilon_2$.}
\EndFor
 \end{algorithmic}
\end{algorithm}

Similar to \cite{ma2023mimo}, the expression of $\tau_m$ is written as
\begin{equation}
\begin{aligned}
\label{deltam}
\tau_m=\sum\limits_{k=1}^K\sum\limits_{j=1}^K\frac{8\pi^2}{\lambda^2}&\left(2\sum\limits_{l\neq m}^M\sum\limits_{p_1=1}^{L_k^t}\sum\limits_{p=1}^{L_k^t}T(p,p_1,k)|[\bW_{j}]_{m,l}|\right.\\
&\left.\quad+[\bW_{j}]_{m, m}\sum\limits_{p_1=1}^{L_k^t}\sum\limits_{p_2=1}^{L_k^t}T(p_2,p_1,k)\right.\\
&\left.\quad+2\sum\limits_{p=1}^{L_k^t}|[\mathbf{f}_k^H]_{p}||[\bar{\bp}_{k,j}]_m|\right).
\end{aligned}
\end{equation}
Then, by ignoring the constant term in \eqref{secondtaylor}, problem $\mathcal{P}_{5.5}$ can be reformulated as 
\begin{subequations}
\begin{align}
\mathcal{P}_{5.6}:\min _{\mathbf{t}_m} 
&\quad \frac{\tau_m}{2} \bt_m^T \bt_m+\left(\nabla \bar{q}\left(\bt_m^{(n)}\right)-\tau_m \bt_m^{(n)}\right)^T \bt_m\label{q2_1}\\
\text { s.t. } 
& \quad{\mathbf t_m} \in \mathcal{S},\label{cons1}\\
&\quad\left\|\mathbf{t}_m-\mathbf{t}_l\right\|_2 \geq D,\quad l=1,\dots,M,\quad l\neq m.\label{cons2} 
\end{align}
\end{subequations}
Note that the objective function \eqref{q2_1} is a convex function. When ignoring the constraints \eqref{cons1} and \eqref{cons2}, the optimal solution of $\bt_m$ is written as
\begin{equation}\label{tmn1}
\bt_{m}^{(n+1)}=-\frac{1}{\tau_m}\nabla \bar{q}\left(\bt_m^{(n)}\right)+\bt_m^{(n)}.
\end{equation}
If $\bt_{m}^{(n+1)}$ satisfies the constraints \eqref{cons1} and \eqref{cons2}, then $\bt_{m}^{(n+1)}$ is the optimal solution of problem $\mathcal{P}_{5.6}$. Otherwise, the first-order Taylor expansion should be exploited to handle the non-convex constraint \eqref{cons2}, which can be written as \cite{ma2023mimo}
\begin{equation}
\begin{aligned}
&\left\|\mathbf{t}_m-\mathbf{t}_l\right\|_2 \\& \geq\left\|\mathbf{t}_m^{(n)}-\mathbf{t}_l\right\|_2+\left(\nabla\left\|\mathbf{t}_m^{(n)}-\mathbf{t}_l\right\|_2\right)^T\left(\mathbf{t}_m-\mathbf{t}_m^{(n)}\right) \\
&=\frac{1}{\left\|\mathbf{t}_m^{(n)}-\mathbf{t}_l\right\|_2}\left(\mathbf{t}_m^{(n)}-\mathbf{t}_l\right)^T\left(\mathbf{t}_m-\mathbf{t}_l\right),
\end{aligned}
\end{equation}
where $\nabla\left\|\mathbf{t}_m^{(n)}-\mathbf{t}_l\right\|_2$ is the gradient vector of $\left\|\mathbf{t}_m^{(n)}-\mathbf{t}_l\right\|_2$ with respect to $\bt_m^{(n)}$. By applying the first-order Taylor expansion to the non-convex constraint \eqref{cons2}, problem $\mathcal{P}_{5.6}$ can be reformulated as
\begin{subequations}
\begin{align}
\mathcal{P}_{5.7}:\min _{\mathbf{t}_m} 
&\quad \frac{\tau_m}{2} \bt_m^T \bt_m+\left(\nabla \bar{q}\left(\bt_m^{(n)}\right)-\tau_m \bt_m^{(n)}\right)^T \bt_m\\
\text { s.t. } 
&\quad{\mathbf{t}_m} \in \mathcal{S},  \\
&\quad \frac{1}{\left\|\mathbf{t}_m^{(n)}-\mathbf{t}_l\right\|_2}\left(\mathbf{t}_m^{(n)}-\mathbf{t}_l\right)^T\left(\mathbf{t}_m-\mathbf{t}_l\right) \geq D, \nonumber\\&\quad\quad\quad\quad\quad\quad\quad\quad\quad\quad l=1,\dots,M,\quad l\neq m.
\end{align}
\end{subequations}
Since problem $\mathcal{P}_{5.7}$ is a convex problem, it can be tackled by exploiting classical convex optimization techniques. The algorithm to tackle problem $\mathcal{P}_{5.4}$ is detailed in \textbf{Algorithm \ref{a_3}}.

\subsection{Outer Layer}\label{ssec:outer}
To make the equality constraints in $\mathcal{P}_{4}$ hold and minimize the SAR value, the penalty factor $\mu$ should be initialized to a small value and gradually increased using the following update
\begin{equation}\label{increaserho}
\mu=\frac{\mu}{a},
\end{equation}
where $0<a<1$ adjusts the precision of the solutions. 

\subsection{Convergence and Complexity Analysis}\label{ssec:convergence}

\begin{algorithm}[t]
\caption{Two-Layer Iterative Algorithm to Handle Problem $\mathcal{P}_{3}$}\label{a_4}
 \begin{algorithmic}[1]
 \State Initialize the thresholds $\varepsilon_3$ and $\varepsilon_4$, the penalty factor $\mu$.
\Repeat
\Repeat  
\State Obtain $\bP$ by optimizing problem $\mathcal{P}_{5.1}$.
\State Obtain $z_{k,j}$ by optimizing problem $\mathcal{P}_{5.2}$.
\State Obtain $\bt$ by optimizing problem $\mathcal{P}_{5.4}$.
\Until{The decrease of the objective value of $\mathcal{P}_{5}$ is below a threshold $\varepsilon_3$.}
\State Update $\mu$ as \eqref{increaserho}.
 \Until{The stopping indicator $\xi$ in \eqref{xi} is below a threshold $\varepsilon_4$.}
 \end{algorithmic}
\end{algorithm}

The two-layer iterative algorithm to tackle $\mathcal{P}_{3}$ is summarized in \textbf{Algorithm \ref{a_4}}, where the stopping indicator is written as
 \begin{equation}
\begin{aligned}\label{xi}
\xi=\sum_{k=1}^K\sum_{j=1}^K|\mathbf h_{k}^H(\bt)\mathbf p_{j}- z_{k,j}|^2.
\end{aligned}
\end{equation}
The convergence of \textbf{Algorithm \ref{a_3}} can be analyzed similar to \cite{ma2023mimo} and is omitted here for simplicity. Then, we analyze the convergence of \textbf{Algorithm \ref{a_4}}. The inner layer is handled by the alternating optimization method, which guarantees the objective values to be non-increasing. Similar to the proof in \cite{7558213}, its convergence is guaranteed. 

The complexity of \textbf{Algorithm \ref{a_4}} is analyzed as follows. The complexity of updating $\bP$ is $\mathcal{O}\left(M^3+K(L_k^t M+M^2)\right)$ while that of optimizing $z_{k,j}$ is $\mathcal{O}\left(K^2\log(1/\varepsilon_5) \right)$, where $\varepsilon_5$ is the accuracy of the bisection search method. The complexity of updating $\bt$ is $\mathcal{O}\left(K^2 M^2 L_k^t\eta_1+M^{2.5}\ln(1/\upsilon)\eta_2 \right)$, where $\eta_1$ is the maximum number of inner iterations in \textbf{Algorithm \ref{a_3}}, $\upsilon$ is the accuracy of the interior-point method, $\eta_2$ is the maximum number of iterations to perform step 8 in \textbf{Algorithm \ref{a_3}}, respectively. Thus, the total computational compexity of \textbf{Algorithm \ref{a_4}} is $\mathcal{O}(\eta_{\rm out}\eta_{\rm inn}(M^3+K(L_k^t M+M^2)+K^2\log_2(1/\varepsilon_5)+K^2 M^2 L_k^t\eta_1+M^{2.5}\ln(1/\upsilon)\eta_2 ))$, where $\eta_{\rm inn}$ and $\eta_{\rm out}$ represent the numbers of inner and outer iterations.

\section{Simulation Results}\label{sec:results}
%
%
In this section, the performance of the proposed SAR-aware FAS design is investigated through numerical results. It has been assumed that the system comprises of a BS with four fluid antennas and four single-FPA users. The number of transmit paths from the BS to the $k$-th user is set as $L_k^t=15$. It is assumed that the azimuth and elevation AoDs are all independently distributed (i.i.d.) variables randomly distributed in $[0,\pi]$ \cite{ma2023mimo}. Assuming that every user is equally distant from the BS, the element of path response vector $\mathbf {f}_k$ is set as a Gaussian distributed variable with zero mean and unit variance. The carrier frequency is set as $30~{\rm GHz}$ and the signal wavelength is $\lambda=0.01~{\rm m}$. The noise variance is set as $\sigma^2=-105~{\rm dBm}$ \cite{lyouee2020}. The minimum distance between the fluid antennas is $D=\lambda/2$ \cite{ma2023mimo}. The given region for the transmit fluid antennas is set as $\mathcal S=[-\lambda,\lambda]\times[-\lambda,\lambda]$.

We suppose that $\gamma_k=\gamma=1$. The SAR budget is set as $Q_0=1.6~{\rm W/kg}$. For four antennas at the BS, the SAR matrix is given by \cite{zhang2019specific}
\begin{equation}
\mathbf{R}=\left[\begin{array}{cccc}
1.6 & -1.2 \jmath & -0.42 & 0 \\
1.2 \jmath & 1.6 & -1.2 \jmath & -0.42 \\
-0.42 & 1.2 \jmath & 1.6 & -1.2 \jmath \\
0 & -0.42 & 1.2 \jmath & 1.6
\end{array}\right].
\end{equation}
The penalty factor is $\mu=10^{-3}$ and the scaling parameter is $a=0.9$. Also, we set $\varepsilon_1=\varepsilon_2=\varepsilon_3=10^{-4}$ and $\varepsilon_4=10^{-7}$.

\begin{figure}[t]
\centering
\includegraphics[scale=0.58]{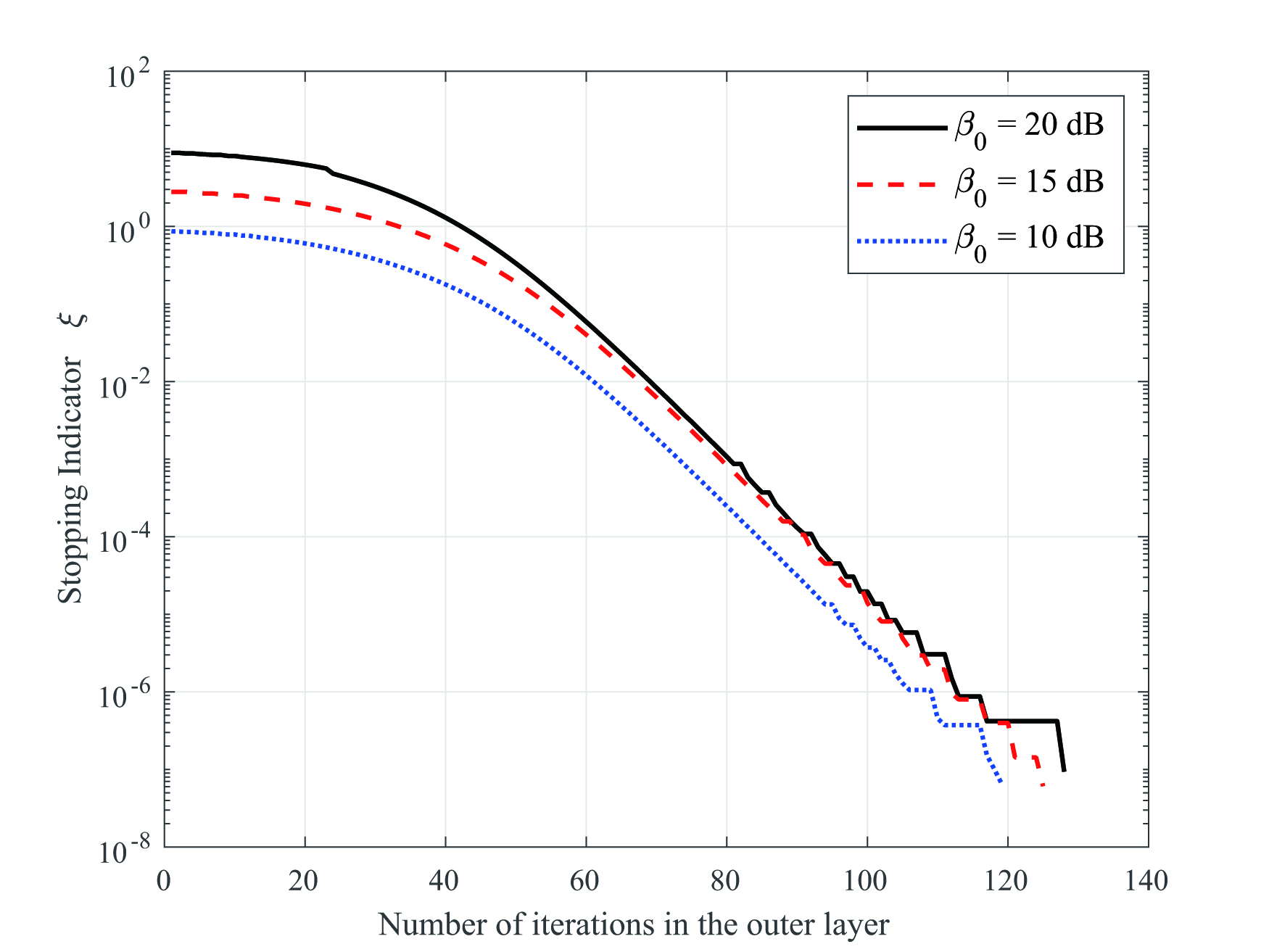}
\captionsetup{font=footnotesize}
\caption{Stopping indicator versus the number of iterations in the outer layer.}\label{fig_penalty}
\end{figure}


We first use the results in Fig.~\ref{fig_penalty} 
to verify the convergence of the proposed algorithm. Fig.~\ref{fig_penalty} shows the stopping indicator of the penalty method versus the total number of iterations in the outer layer. We observe that after about $120$ iterations, the stopping indicator $\xi$ can be met with the threshold $\varepsilon_4$, meaning that the equality constraints can always be satisfied. Also, as the SINR threshold $\beta_0$ rises, the number of iterations increases. 


To show the effectiveness of the SAR constraint in $\mathcal{P}_1$, we compare the proposed scheme with the without SAR scheme and the adaptive backoff scheme:
\begin{enumerate}
\item[1)] \textbf{Without SAR}: Consider the minimum weighted SINR maximization problem with a power constraint instead of the SAR constraint, i.e.,
\begin{equation}
\begin{aligned}
\mathcal{P}_6: \max _{\bP,\bt}
&\quad\min_{k} \frac{1}{\gamma_k}\operatorname{SINR}_k(\bP,\bt) \\
\text { s. t. }
& \quad\sum_k||\bp_k||^2 \leq {P}_t,\\
& \quad{\mathbf t} \in \mathcal{S},\\
&\quad\left\|\mathbf{t}_m-\mathbf{t}_l\right\|_2 \geq D,\quad 1\leq m\neq l\leq M.
\end{aligned}
\end{equation}
Here, the transmit power budget is set as $P_t=2~{\rm W}$.
\item[2)] \textbf{Adaptive Backoff} \cite{ying2015closed}: First consider problem $\mathcal{P}_6$ and assume that the solution of the precoding matrix is $\bar\bP$ and the solution of the transmit fluid antenna positions is $\bar\bt$. Then, the SAR constraint is satisfied by multiplying the precoded matrix by a backoff factor $\alpha$, i.e.,
\begin{equation}
\bP_{\mathrm{opt}}=\alpha\bar\bP,
\end{equation}
where
\begin{equation}
\alpha=\min \left\{1, \frac{Q_{0}}{\sum_{k=1}^K \bar\bp_k^H\bR\bar\bp_k}\right\}.
\end{equation}
\end{enumerate}

\begin{figure}[t]
\centering
\includegraphics[scale=0.58]{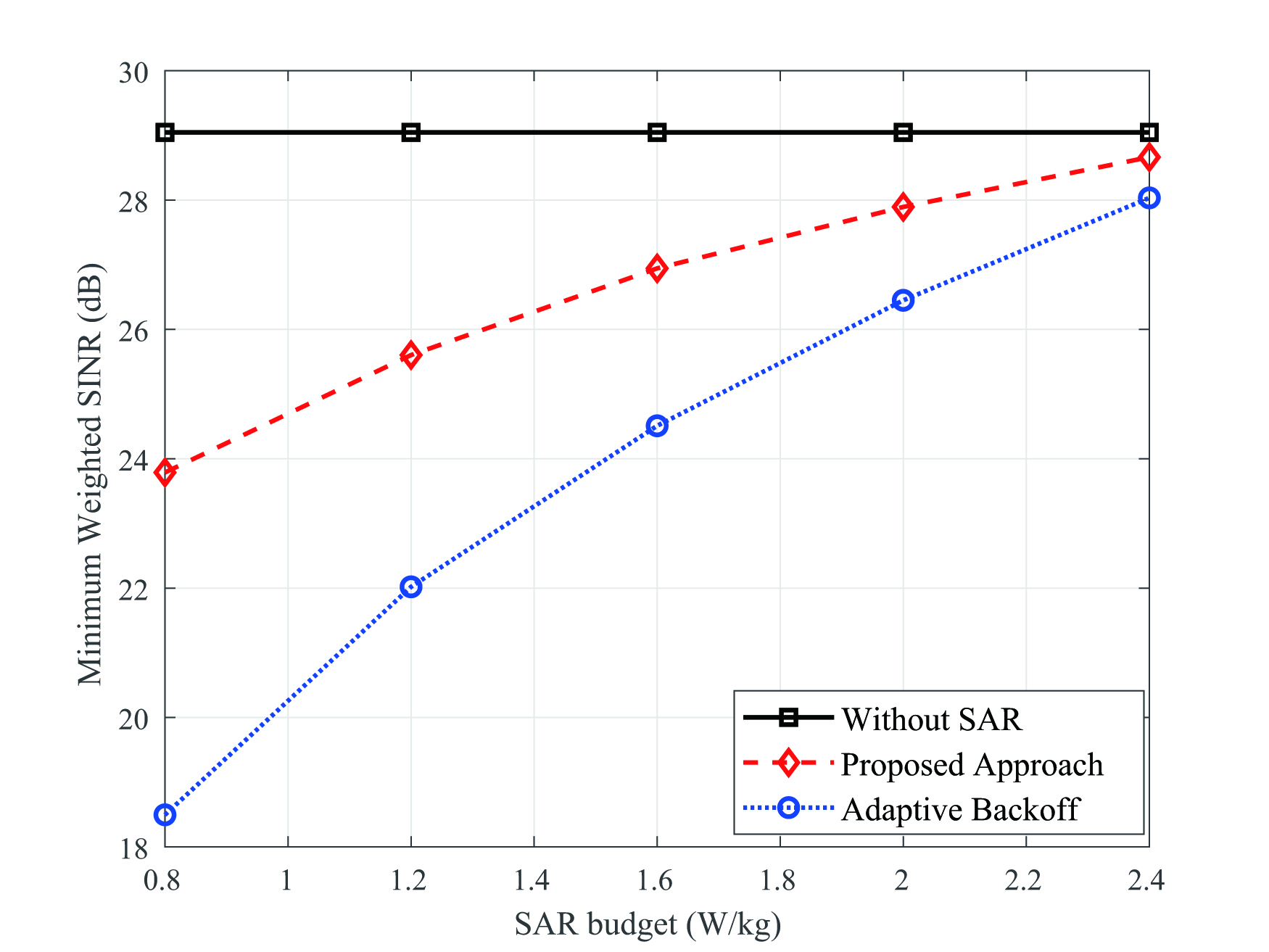}
\captionsetup{font=footnotesize}
\caption{Minimum weighted SINR versus the SAR budget.}\label{fig_backoff}
\end{figure}

Fig.~\ref{fig_backoff} compares the proposed scheme with baselines, i.e., the without SAR and adaptive backoff schemes for the optimization problem $\mathcal{P}_1$. We see that the proposed approach greatly outperforms the adaptive backoff scheme regarding the minimum weighted SINR. The gap between the proposed approach and adaptive backoff becomes smaller as the SAR budget rises. In fact, as the SAR budget rises, the backoff factor gradually increases to one, which means that the solution of the precoding matrix in these two schemes becomes closer. In addition, the curves according to without SAR and proposed approaches become closer with the increase of the SAR budget. When the SAR budget $Q_0$ is relatively small, the SAR constraint is the main impact of the minimum weighted SINR. When the SAR budget becomes relatively large, the power constraint starts to dominate the system's minimum weighted SINR.

To illustrate the advantages of FAS, we compare the proposed approach with the following baselines:
\begin{enumerate}
\item[1)] \textbf{Alternating position selection (APS)}: The transmit region $\mathcal S$ is quantized into discrete positions. The spacing between each position is $D=\lambda/2$. The optimization problem is tackled by an exhaustive search among the discrete locations.
\item[2)] \textbf{FPA}: The BS are equipped with fixed-postion uniformly linear arrays. The spacing between antenna elements is $\lambda/2$.
\end{enumerate}

\begin{figure}[t]
\centering
\includegraphics[scale=0.58]{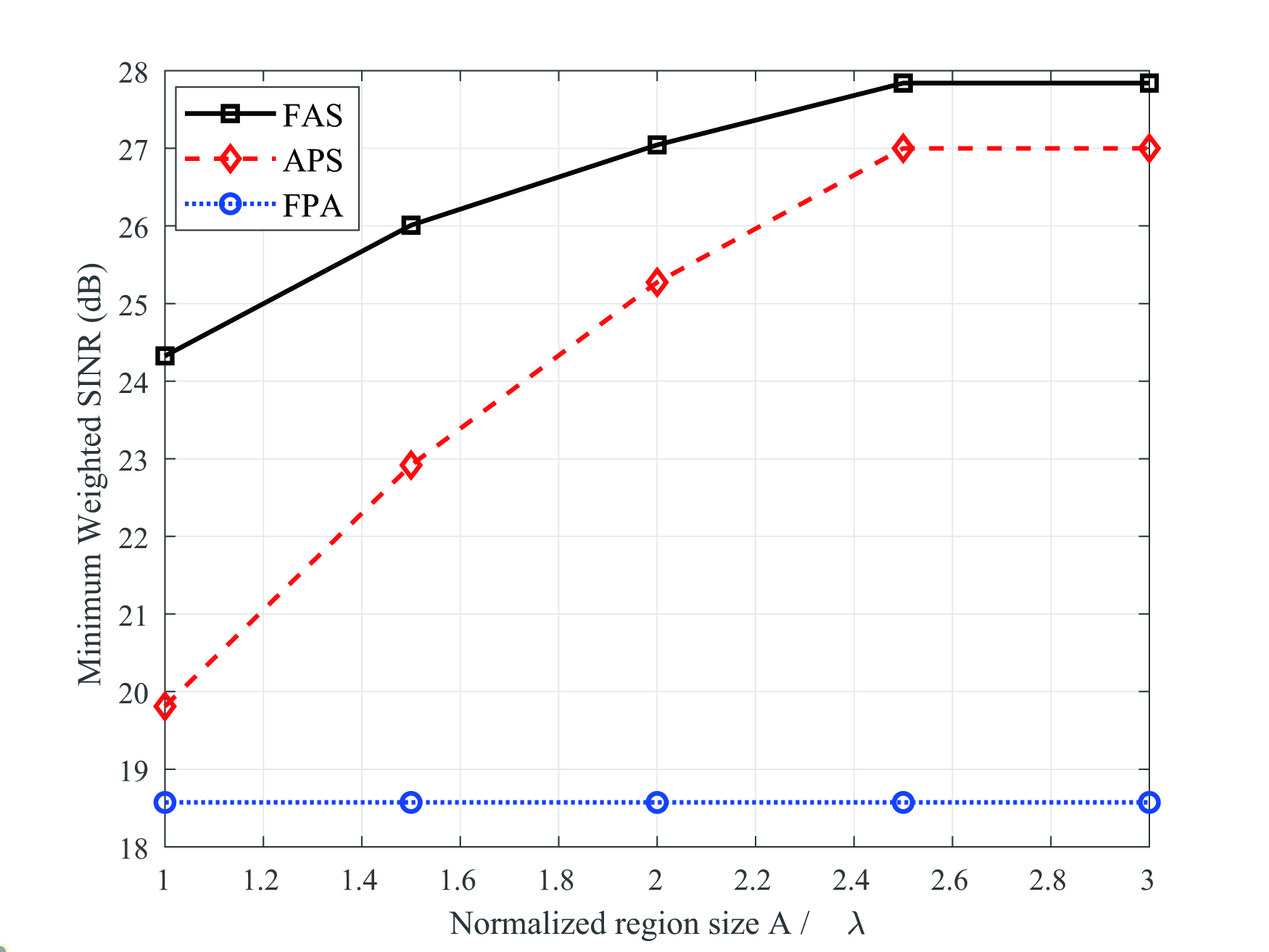}
\captionsetup{font=footnotesize}
\caption{Minimum weighted SINR versus the size of the normalized region.}\label{fig_FA_A}
\end{figure}

\begin{figure}[!t]
\centering
\subfloat[$L_k^t=15$]{
\includegraphics[scale=0.58]{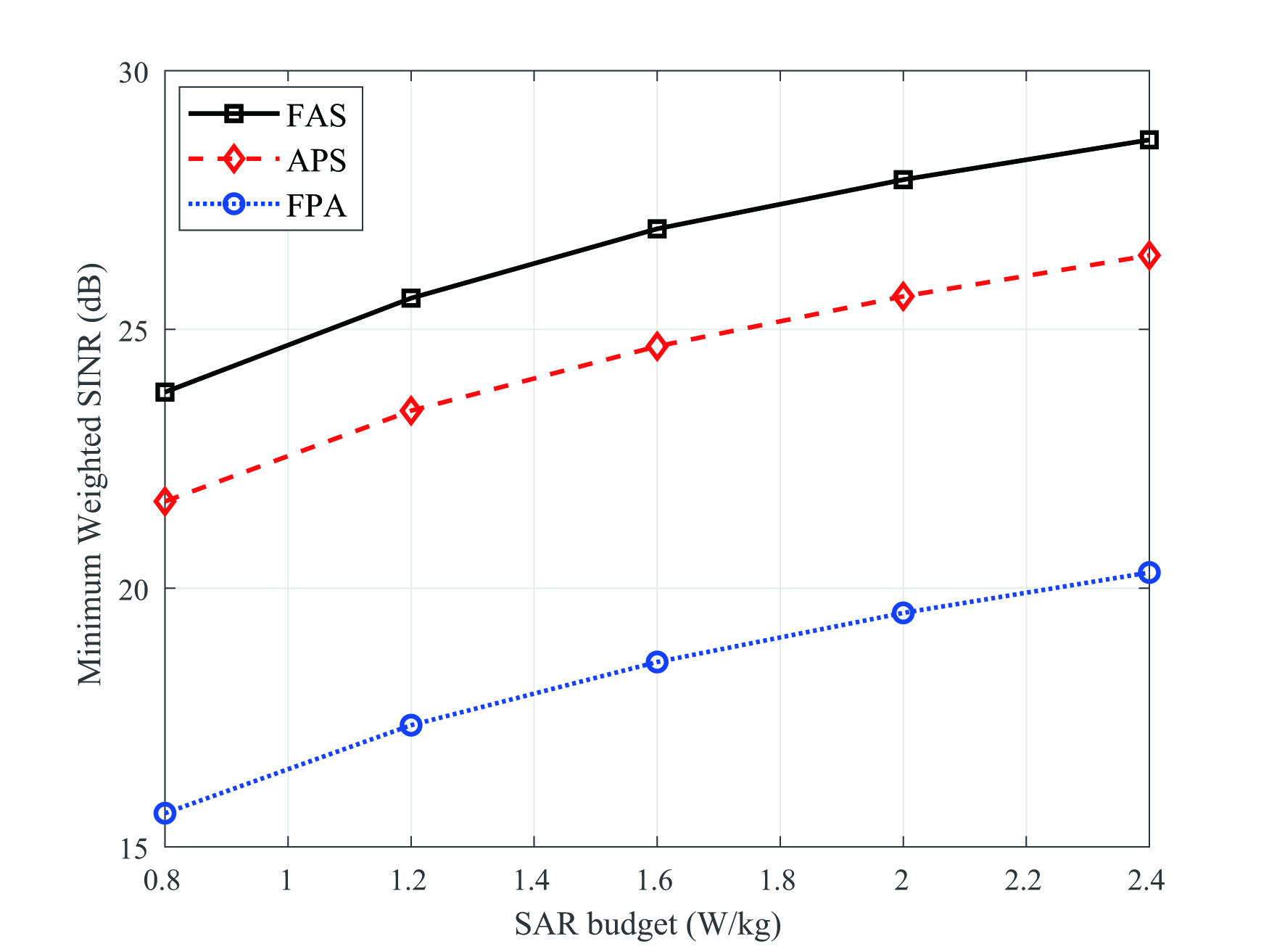}}\\
\vspace{-3mm}
\subfloat[$L_k^t=5$]{
\includegraphics[scale=0.58]{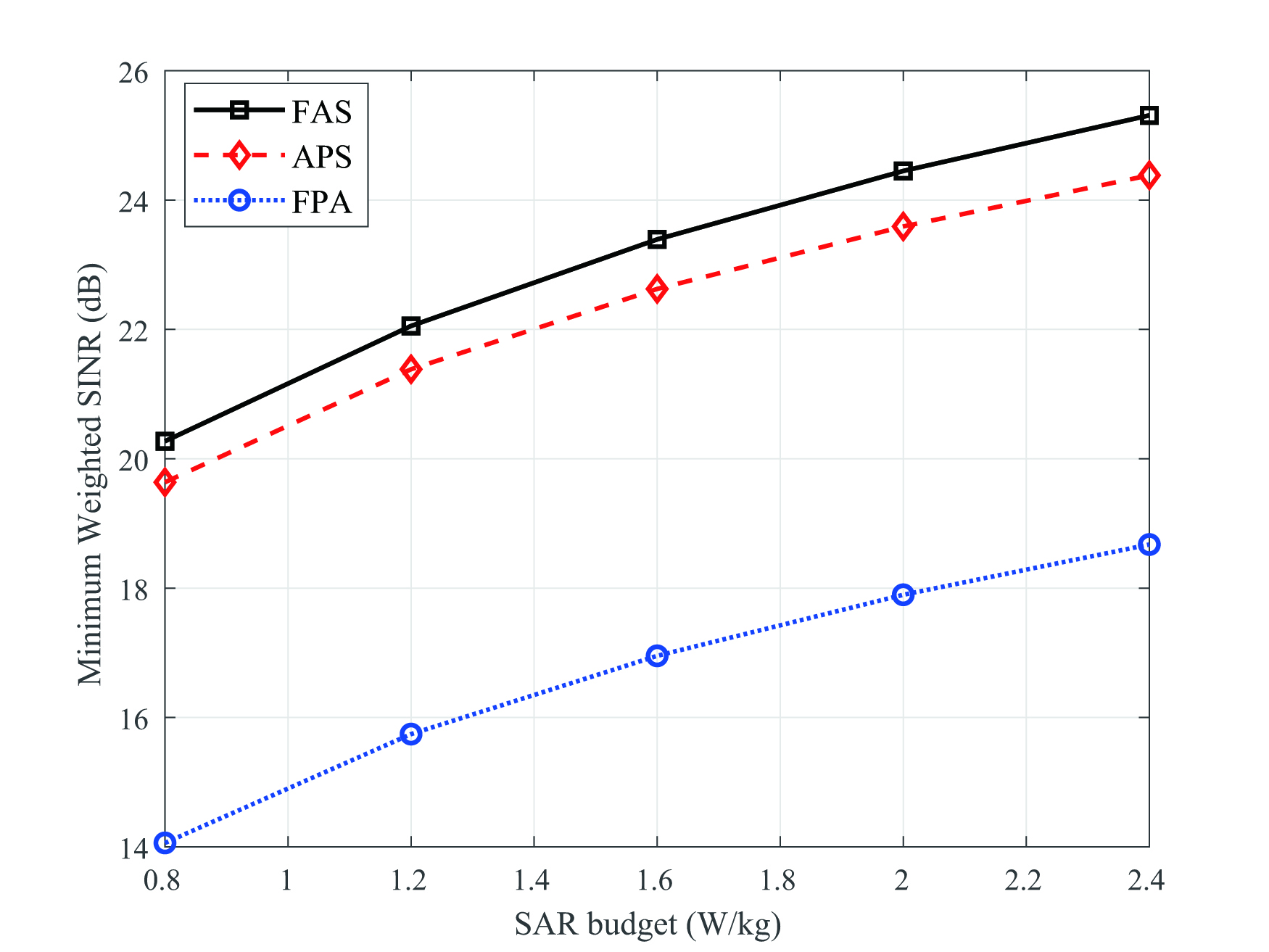}}
\captionsetup{font=footnotesize}
\caption{Minimum weighted SINR with respect to SAR budget for different number of paths.}\label{fig_FA_SAR_L}
\end{figure}

Fig.~\ref{fig_FA_A} compares the proposed FAS design with APS and FPA designs in terms of the minimum weighted SINR for the optimization problem $\mathcal{P}_1$. We observe that the FAS design can achieve substantial gains over the APS and FPA designs. This is due to the fact that fluid antennas can switch to any position in the given region to obtain a larger minimum weighted SINR. In other words, FAS can utilize extra DoFs to obtain a more favourable channel condition. The upward trend of the FAS curve represents that the increase of the transmit fluid antenna region provides more DoF and allows the fluid antenna to move freely to more desirable positions. Additionally, the curve corresponding to the FAS design is saturated when $A=2.5\lambda$, which means that a larger transmit region is no longer needed.

Fig.~\ref{fig_FA_SAR_L} compares the minimum weighted SINR with other baselines regarding the SAR budget under a different number of paths for the optimization problem $\mathcal{P}_1$. We can observe that the minimum weighted SINR value of these three designs shows an increasing tendency as the SAR budget increases, and the proposed FAS design consistently outperforms the APS and FPA designs. By comparing the results in Figs.~\ref{fig_FA_SAR_L}(a) and \ref{fig_FA_SAR_L}(b), it can be observed that the minimum weighted SINR of all curves increases as the number of paths rises. In addition, it is clearly shown that the gap between FAS and APS and FPA designs becomes wider with the increasing number of paths. This is because increasing the number of paths leads to higher multipath diversity gain and stronger small-scale fading, which gives more DoFs for FAS design to further increase the minimum weighted SINR.

\begin{figure}[t]
\centering
\includegraphics[scale=0.58]{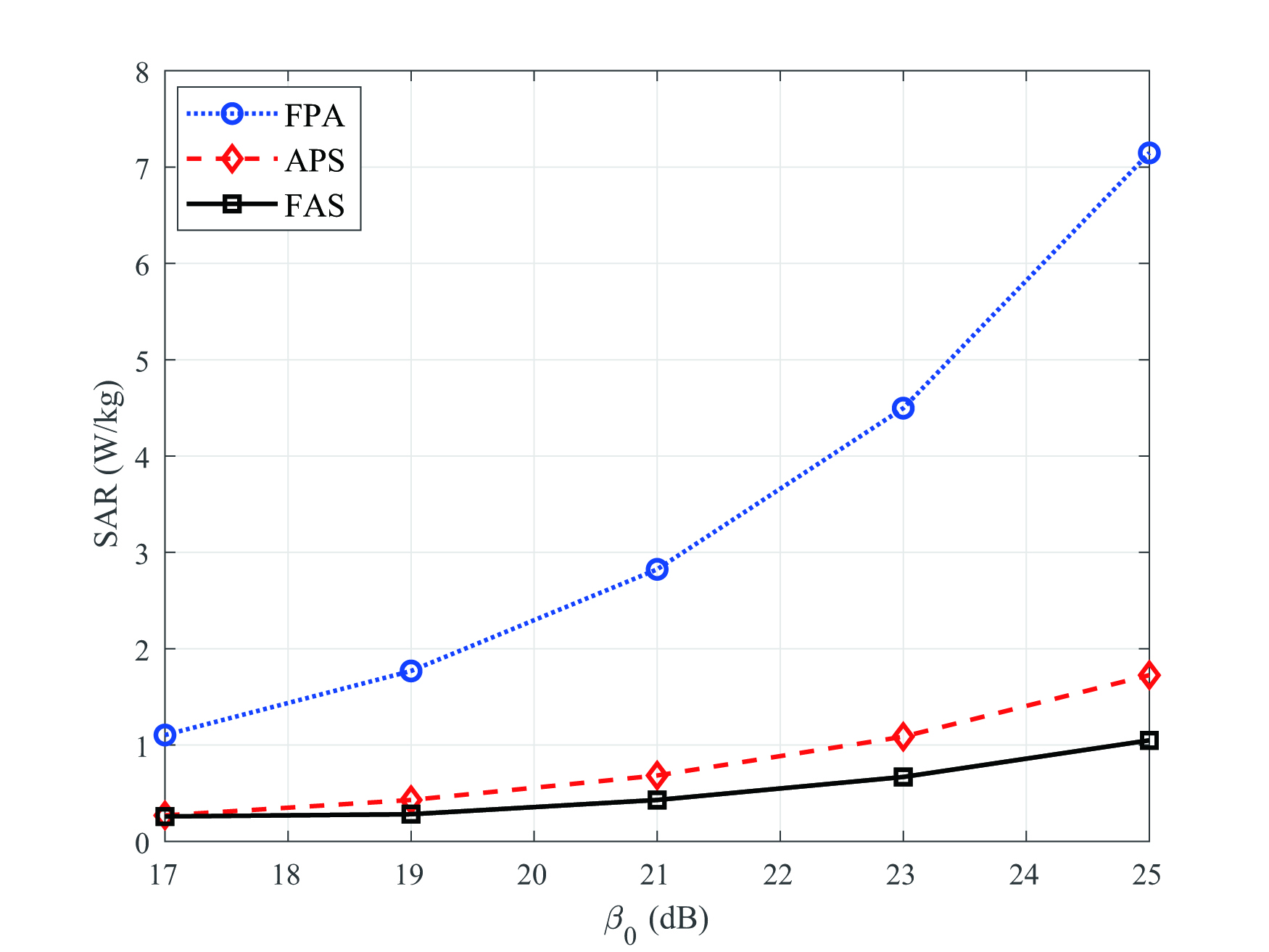}
\captionsetup{font=footnotesize}
\caption{SAR with respect to $\beta_0$.}\label{fig_beta_SAR}
\end{figure}

\begin{figure}[t]
\centering
\includegraphics[scale=0.58]{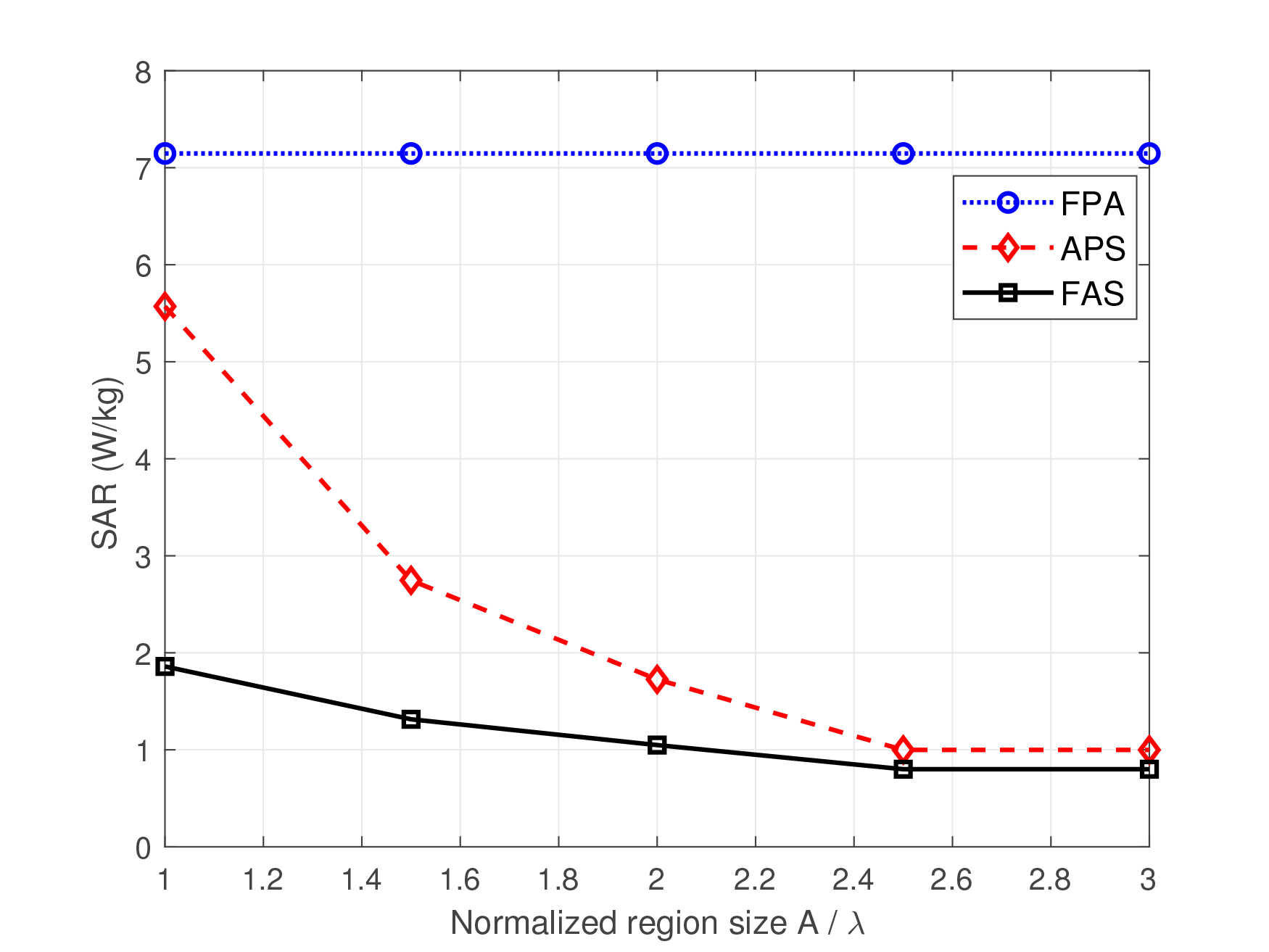}
\captionsetup{font=footnotesize}
\caption{SAR with regard to the size of the normalized region.}\label{fig_A_SAR}
\end{figure}

Fig.~\ref{fig_beta_SAR} illustrates SAR with respect to $\beta_0$ for the optimization problem $\mathcal{P}_3$. We can observe that as the SINR threshold $\beta_0$ rises, the SAR value shows an upward tendency, which means that a larger SAR value is needed to satisfy the stricter SINR constraints. Moreover, it is clear that the FAS design can reduce SAR value compared with FPA and APS designs, which confirms the effectiveness of the SAR-aware FAS design.

Fig.~\ref{fig_A_SAR} shows the SAR value against the size of the normalized transmit region for the optimization problem $\mathcal{P}_3$. It can be seen that the proposed FAS design yields a smaller SAR value compared with the FPA and APS designs. In fact, FAS can adaptively reconfigure the channel into a more desirable state by adjusting the antenna positions. In addition, as the transmit region size increases, the SAR value decreases and is finally saturated when $A=2.5\lambda$.

\section{Conclusion}\label{sec:conclude} 
This paper investigated the SAR-aware multiuser MIMO downlink communications assisted by FAS. Two different optimization targets were considered, namely, (1) SAR minimization and (2) SINR balancing. We first proposed a two-layer iterative algorithm to minimize the SAR under the SINR and FAS constraints. Then, an algorithm was proposed to tackle the minimum weighted SINR maximization problem by studying its relationship with the SAR minimization problem. Numerical results demonstrated the superior performance of the proposed design in comparison to adaptive backoff and FPA designs.

%

\bibliographystyle{IEEEtran}

\end{document}